\documentclass[12pt,a4paper]{article}

\makeatletter
\def\@cite#1#2{$^{\hbox{\scriptsize{#1\if@tempswa , #2\fi})}}$}
\def\thebibliography#1{
 \section*{References\@mkboth{References}{References}}
 \list{\arabic{enumi})}
   {\settowidth{\labelwidth}{[#1]}
    \leftmargin=\labelwidth
    \advance \leftmargin by \labelsep
    \usecounter{enumi}}
 \def\newblock{\hskip .11em plus .33em minus .07em}
 \sloppy
 \clubpenalty=4000 \widowpenalty=4000
 \sfcode`\.=1000\relax}
        
\renewcommand{\title}[1]{
\begin{center} \Large \bf #1 \end{center}
}

\renewcommand{\author}[3]{
 \begin{center} #1 \\
  {\it #2} \\
  {\small E-mail: \texttt{#3}}
 \end{center}
\addvspace{\baselineskip}
}

\usepackage{amssymb}
\usepackage{amsmath}

\usepackage{theorem}
\theoremstyle{break}

\theoremstyle{break}

\theoremstyle{break}

\font\mybb=msbm10 at 12pt
\def\bb#1{\hbox{\mybb#1}}

\font\mybb=msbm10 at 12pt

\def\ad{\hbox{\rm ad}}

\def\half{{1\over 2}} 

\begin{document}

%%%%%%%%%%title%%%%%%%%%%%%%%%%%%%
\begin{titlepage}

\title{Partition Functions of Supersymmetric\\
 Gauge Theories in Noncommutative ${\mathbb R}^{2D}$ 
\\
and their Unified Perspective }

\author{Akifumi Sako${}^{\dagger}$ \ , 
\ Toshiya Suzuki${}^{\ast} $  \\ \ }
{${}^{\dagger}$ Department of Mathematics, Faculty of Science and
 Technology, Keio University\\
3-14-1 Hiyoshi, Kohoku-ku, Yokohama 223-8522, Japan\\ \ \\
${}^{\ast}$ Department of Physics, Faculty of Science, Ochanomizu University\\
2-1-1 Otsuka, Bunkyo-ku, Tokyo 112-8610, Japan\\ \  }
{${}^{\dagger}$ sako@math.keio.ac.jp\\
\makebox{}\hspace{18mm} ${}^{\ast}$ tsuzuki@phys.ocha.ac.jp}

\vspace{1cm}

\abstract{
We investigate 
cohomological gauge theories in noncommutative ${\mathbb R}^{2D}$.
We show that vacuum expectation values of the theories 
do not depend on noncommutative parameters, and
the large noncommutative parameter limit 
is equivalent to the dimensional reduction.
As a result of these facts, we show that a partition function of
a cohomological theory defined in noncommutative
${\mathbb R}^{2D}$ and a partition function of a cohomological
field theory in ${\mathbb R}^{2D+2}$
are equivalent if they are connected through dimensional
reduction.
Therefore, we find several partition functions of supersymmetric
gauge theories in various dimensions are equivalent.
Using this technique, we determine the 
partition function of the ${\cal N}=4$ U(1) gauge theory 
in noncommutative ${\mathbb R}^4$, where its action
does not include a topological term.
The result is common among 
(8-dim , ${{\cal N}=2}$), (6-dim , ${{\cal N}=2}$),
(2-dim , ${{\cal N}=8}$) and the IKKT matrix model given 
by their dimensional reduction
to 0-dim.
%We also discuss the partition function with the topological term.
}

\baselineskip 5mm

\end{titlepage}

%%%%%%%%%%%%%%%%%%%%%%%%%%%%%%%%%%%%%%%%%%%%%%%%%%%%%%%%%%%%%%%%%%%%%%
%%%%%%%%%%%%%%%%% N=4 Yang-Mills case %%%%%%%%%%%%%%%%%%%%%%%%%%%%%%%%
%%%%%%%%%%%%%%%%%%%%%%%%%%%%%%%%%%%%%%%%%%%%%%%%%%%%%%%%%%%%%%%%%%%%
\section{Introduction}\label{intro}
The first break through of the recent calculation technology for
${\cal N}= 2$ supersymmetric Yang-Mills theories 
is brought by Nekrasov \cite{Nekrasov4,Nekrasov-Okounkov}.
After \cite{Nekrasov4}, many kinds of developments appear
in ${\cal N} \ge 2$ supersymmetric Yang-Mills theories and
string theories corresponding to them.
{}From those analysis, it is found that different dimension theories 
are related each other 
\cite{Eguchi-Kanno,Maeda-Nakatsu-Takasaki-Tamakoshi,Matsuo-Matsuura-Ohta,Tachikawa,Wijnholt}.
There is more example that the different dimensional theories
are connected to each other.
For example, Dijkgraaf and Vafa show that
some correlation functions in
matrix theories and ${\cal N}=1$ Yang-Mills theories 
are equivalent 
%satisfy
%the same Schwinger-Dyson equations
\cite{dijkgraaf-Vafa}.
It goes on and on.
These facts imply the existence of some kind of unified perspectives.
One of the ideas to explain the unification 
is the 'tHooft's large $N$ gauge theory and 
string correspondence.
Until now, many investigations from this point of view
are reported.
Meanwhile, the large $N$ gauge theories are similar to noncommutative
theories in the operator formalism in some infinite dimensional
Hilbert space with discrete basis.
In this article, we suggest a simple way to understand
the reason why partition functions of various dimensional 
supersymmetric gauge theories
are given as same form or have relations with each other.
The basic idea of the way is given in \cite{sako,sako1,sako2}.
Cohomological gauge theories in Euclidian spaces are invariant under the 
noncommutative parameter shifting, as we will see it in the next section.
When we take the large noncommutative parameter limit,
kinetic terms become irrelevant like dimensional reduction, 
then the partition function is essentially computable 
by using lower dimensional theories.
{}From this fact, we will explain that partition functions 
in various dimensions are equivalent.\\

%%%%%%%%%% 
Here is the organization of this article.
In section \ref{ncym}, invariance of cohomological
field theories in noncommutative ${\mathbb R}^{2D}$
(N.C. ${\mathbb R}^{2D}$ for short )  under deformation of 
noncommutative parameters will be
proved formally.
This invariance is not usual symmetry, because the action is deformed.
Nevertheless, expectation values and partition functions are invariant.
Particularly, we will treat the 
${\cal N}=2$ and ${\cal N}=4$ Yang-Mills theories 
in N.C. ${\mathbb R}^{4}$
as examples.
In section \ref{upf}, 
universality of the partition functions will be investigated.
By using the result of section \ref{ncym}, we will show that the 
several partition functions in
different dimensions are equivalent.
%%%%%%%%%%%%% REV5
(In appendix \ref{B}, concrete discussions
for some models will be given again.)
%%%%%%%%%%%%%
In section \ref{pfn4}, by the technique of section \ref{ncym}
we will calculate the partition function of the
${\cal N}=4$ U(1) gauge theory in N.C. ${\mathbb R}^4$
without the terms proportional to the instanton number $\int F\wedge F$.
This partition function is equal to  partition 
functions of several dimensions.
In section \ref{modulispace}, the moduli space of 
${\cal N}=4$ U(1) gauge theory in N.C. ${\mathbb R}^4$
will be discussed.
The partition function of ${\cal N}=4$ U(1) theory with
$\int F\wedge F$ will be investigated, too.
In section \ref{conclusion}, we will summarize this article.

%%%%%%%%%%%%%%%%%%%%%%%%%%%%%%%%%%%%%%%%%%%%%%%%%%%%%%%%
%%%%%%%%%%%%%%%%%%%%%%%%%%%%%%%%%%%%%%

\section{ N.C. Cohomological Yang-Mills Theory}\label{ncym}
In this section, we investigate some important
properties of the cohomological Yang-Mills theories
in N.C. ${\mathbb R}^{2D}$
whose noncommutativity is defined as
\begin{eqnarray}
[x^{\mu} , x^{\nu} ] = i \theta^{\mu \nu} \ ,
\end{eqnarray}
where the $\theta^{\mu \nu}$ is an element of an antisymmetric matrix
and called noncommutative parameter.\\

Since action functionals of cohomological field theories
are defined by  BRS-exact
functionals like $\hat{\delta} \Psi[\phi_i ]$, where $\hat{\delta}$
is a some BRS operator and $\{ \phi_i \}$ represent all considered fields 
and
$\Psi$ is a some fermionic functional,
the partition function of the cohomological field theory
is invariant under any infinitesimal
variation $\delta'$ which commutes
(or anti-commutes) with the BRS transformation:
%%%%%%%%%%%%%%%
\begin{eqnarray}
\label{2.13}
\hat{\delta} \delta' &=& \pm \delta' \hat{\delta}, \nonumber\\
\delta' \ Z_{\theta} &=& \int \prod_i {\cal D}\phi_i
%{\cal D} \psi {\cal D}\chi {\cal D}H
\ \ \delta' \left( -\int dx^{2D} \hat{\delta} \Psi \right) \ \exp 
\left( -S_\theta
\right)
\nonumber\\
&=& \pm \int \prod_i {\cal D}\phi_i
%{\cal D} \psi {\cal D}\chi {\cal D}H
\ \
\hat{\delta} \left( -\int dx^{2D} \delta' \Psi \right) \
\exp \left( -S_\theta \right)
=0.
\end{eqnarray}
Let $\delta_{\theta}$ be the infinitesimal deformation
operator of
the noncommutative parameter $\theta$ which operates as
\begin{eqnarray}
\delta_{\theta}\ \theta^{\mu \nu} = \delta \theta^{\mu \nu},
\end{eqnarray}
where $\delta \theta^{\mu \nu}$ are
some infinitesimal anti-symmetric two form
elements.
If $\delta_{\theta}$ and BRS operator $\hat{\delta}$ commute
each other, then the partition function is invariant.
Indeed, there is some examples such that
$\hat{\delta} \delta_{\theta} = \delta_{\theta} \hat{\delta}$,
and partition functions are calculated by using this property
\cite{sako,sako1,sako2}.
\\

In this article, cohomological Yang-Mills theories in noncommutative
Euclidian spaces are discussed.
If there is a gauge symmetry, the BRS-like transformation is slightly
different from the one of non-gauge theory.
The BRS-like symmetry is not nilpotent but
\begin{eqnarray}
\hat{\delta}^2 =\delta_{g, \theta}, \label{del_g}
\end{eqnarray}
where
$\delta_{g, \theta}$ is a gauge transformation operator
deformed by some noncommutative deformation method like the star product
$\mbox{\Large $*$}_{\theta}$.
As occasion arises, the gauge transformation $\delta_{g, \theta}$
is defined as one including global symmetry transformations.
The partition function of the noncommutative cohomological
field theory is invariant
under changing noncommutative parameters when the BRS transformation
does not depend on the noncommutative parameters,
because the BRS transformation $\hat{\delta}$ and the
$\theta$ deformation $\delta_{\theta}$ commute.
Conversely, when the definition of the BRS-like operator (\ref{del_g}) 
depends on
the noncommutative parameter $\theta$, then $\hat{\delta}$ and
$\delta_{\theta}$ do not commute :
\begin{eqnarray}
\delta_{\theta} \hat{\delta} \neq \hat{\delta} \delta_{\theta} \Rightarrow
\delta_{\theta} \hat{\delta} = \hat{\delta}' \delta_{\theta} ,
\end{eqnarray}
where $\hat{\delta}'$ is a BRS-like operator that generates
the same transformations as the original BRS-like operator $\hat{\delta}$
, except for the square. The square of $\hat{\delta}$ is defined by
\begin{eqnarray}
\hat{\delta}'{}^2 =\delta_{g, \theta+\delta \theta}.
\end{eqnarray}
Since the gauge symmetry is defined by using noncommutative parameter
$\theta^{\mu \nu}+\delta\theta^{\mu \nu}$ after the $\delta_{\theta}$ 
operation,
this difference arises.
This fact makes a little complex problem to prove the $\theta$-shift
invariance of noncommutative
cohomological Yang-Mills theory in comparison with
the case of non-gauge theory.\\

Note that the essential point of this problem is not nilpotent property
changing,
but $\theta$ dependence of the definition of the BRS operator.
(In fact, we can construct the BRS operator for the cohomological Yang-Mills
theory as a nilpotent operator \cite{P.R}. )\\

%%%%%%%%%%%%%%%%%%%%%%
However, we can prove the invariance of the partition function
of cohomological Yang-Mills theory in N.C. ${\mathbb R}^{2D}$
under the noncommutative parameter deformation.
For simplicity, we take
$$(\theta^{\mu \nu}) = \bigoplus_i \epsilon^{2i-1 , 2i } \theta =
\theta \left(
\begin{array}{cc}
0 & 1 \\
-1 & 0
\end{array}
\right) \oplus \cdots \oplus
\left(
\begin{array}{cc}
0 & 1 \\
-1 & 0
\end{array}
\right) \ ,
$$
where $ \epsilon^{2i-1 , 2i }$ is an antisymmetric tensor such that
$\epsilon^{2i-1 , 2i }= -\epsilon^{2i , 2i-1 }=1$,
and we restrict the $\theta$ deformation to
$$\theta \rightarrow \theta+\delta \theta \ \ .
$$

In the following, we only use operator formalisms to describe the
noncommutative field theory, therefore
the fields are operators acting on the Hilbert space ${\cal H}$.
Then differential operators $\partial_{\mu}$ are expressed by using
commutation brackets
$ -i \theta^{ -1}_{\mu \nu}[ x^{\nu} , \ *] \equiv [ \hat{\partial}_{\mu} , 
* ]$
and $\int d^{2D} x $ is replaced with $ det (\theta )^{1/2}Tr_{\cal H} $.
Therefore the noncommutative parameter deformation is
equivalent with replacing $ -i \theta^{ -1}_{\mu \nu}[ x^{\nu} , \ ] $
and $ det (\theta )^{1/2}Tr_{\cal H} $ by
$ -i (\theta+\delta \theta)^{ -1}_{\mu \nu}[ x^{\nu} , \ ] $
and $ det (\theta +\delta \theta)^{1/2}Tr_{\cal H} $, respectively.\\

Let us consider
Donaldson-Witten theory
(topological twisted ${\mathcal N}=2$ Yang-Mills theory) on
N.C. ${\mathbb R}^4$ \cite{Witten}.
This theory is constructed by bosonic fields
$(A_{\mu}, H_{\mu \nu}^+ , \bar{\phi},  \phi )$ and
fermionic fields $(\psi_{\mu} , \chi_{\mu \nu}^+ , \eta )$
, where $(A_{\mu}, H_{\mu \nu}^+ , \bar{\phi} )$
and $( \psi_{\mu} ,  \chi_{\mu \nu}^+ , \eta )$ are supersymmetric
(BRS) pairs,
\begin{eqnarray}
\chi_{\mu\nu}^+, \; H_{\mu\nu}^+ \; &\in &\; \Omega^{2,+}({\mathbb R}^4 ,\ad 
P),
\;\;\;\;\;
\psi_\mu \; \in \; \Omega^1({\mathbb R}^4 ,\ad P),
\nonumber \\
\eta, \; \bar{\phi}, \; \phi \; &\in &\; \Omega^{0}({\mathbb R}^4 ,\ad P).
\label{cveintesiete}
\end{eqnarray}
Their ghost numbers are assigned as
($A_\mu$,$\chi_{\mu\nu}^+$,$H_{\mu\nu}^+$,$\psi_\mu$,$\eta$,
$\bar{\phi}$,$\phi$)=($0$,$-1$,$0$,$1$,$-1$,$-2$,$2$).
The BRS-like operator is defined by
\begin{eqnarray}
\hat{\delta} A_\mu &= \psi_\mu,  \mbox{\hskip2cm}
\hat{\delta} \chi_{\mu\nu}^+ &=  H_{\mu\nu}^+,
\nonumber \\
\hat{\delta} \psi_{\mu} &= D_{\mu} \phi, \mbox{\hskip1.5cm}
\hat{\delta} H_{\mu\nu}^+ &= i[\chi_{\mu\nu}^+,\phi], \nonumber \\
\hat{\delta} \phi &= 0, \mbox{\hskip2.3cm}
\hat{\delta} \bar{\phi} &= \eta,
\mbox{\hskip2cm}
\hat{\delta}\eta = i [\bar{\phi},\phi],
\label{N=2BRS}
\end{eqnarray}
where the covariant derivative is defined by
$D_{\mu}\  * := [\hat{\partial}_{\mu} + i A_{\mu}\ ,\  * \ ]$
with $\hat{\partial}_{\mu}:= -i \theta^{ -1}_{\mu \nu} x^{\nu} $.
When we consider only the case of N.C. ${\mathbb R}^{2D}$,
field theories are expressed by the Fock space formalism.
(See appendix \ref{fock}.)
In the Fock space representation,
fields are expressed as $A_{\mu}= \sum {A_{\mu}}^{n_1 n_2}_{m_1 m_2}
|n_1 ,n_2 \rangle \langle m_1 , m_2|$ ,
$\psi_{\mu}= \sum {\psi_{\mu}}^{n_1 n_2}_{m_1 m_2}
|n_1 ,n_2 \rangle \langle m_1 , m_2|$ , etc.
Therefore, the above BRS transformations are expressed as
\begin{eqnarray}
\hat{\delta} {A_{\mu}}^{n_1 n_2}_{m_1 m_2} = {\psi_{\mu}}^{n_1 n_2}_{m_1 
m_2}
\ \ , \cdots \ \ .
\end{eqnarray}
Let us define gauge fermions as
\begin{eqnarray}
&&\Psi = Tr_{\cal H}  tr\left[2\chi_{\mu\nu}^+(
-iF^{+\mu\nu}+\half H^{+ \mu\nu})\right],\nonumber\\
&&\Psi_{\rm  proj}= -Tr_{\cal H}  tr\left[\bar{\phi}  D_\mu\psi^\mu
\right],
\label{N=2gaugefermion}
\end{eqnarray}
then the action functional is given by
\begin{eqnarray}
S&=& Tr_{\cal H} \
L ( A_{\mu}, \dots ;
\hat{\partial}_{z_i}, \hat{\partial}_{\bar{z}_i} ) \nonumber \\
&=& Tr_{\cal H}  tr
 \hat{\delta}(\Psi +\Psi_{\rm  proj})  \label{ctreinta} \\
&=& Tr_{\cal H}  tr
\, \Big({F^+}^2-4i\chi^{+\mu\nu} D_\mu \psi_\nu -\eta D_\mu
\psi^\mu +i \phi \{ \chi_{\mu\nu}^+,\chi^{+\mu\nu} \} \nonumber
\\ &&\,\,\,\,\,\,\,\,\,\,\,\,\,\,\,\,\,\,\,\,\,\,\,\,\,\,
\,\,\,\,\,\,\,\,\,\,\,\,\,
-i \bar{\phi} \{\psi_\mu,\psi^\mu\}
-\bar{\phi} D_\mu D^\mu\phi \Big)\ \ , \nonumber
\end{eqnarray}
where $tr$ is trace for gauge group.
In this article, we omit to note $det(\theta )^{1/2}$
that is an overall factor,
for economy of space.
Let us change the dynamical variables as
\begin{eqnarray}
&&A_{\mu} \rightarrow \frac{1}{\sqrt{\theta}} \tilde{A}_{\mu} , \ \
\psi_{\mu} \rightarrow \frac{1}{\sqrt{\theta}} \tilde{\psi}_{\mu}  , \ \
\bar{\phi} \rightarrow \frac{1}{\theta} \tilde{\bar{\phi}} , \ \
\eta \rightarrow \frac{1}{\theta} \tilde{\eta}
\nonumber \\
&&
\chi_{\mu \nu}^+ \rightarrow \frac{1}{\theta} \tilde{\chi}_{\mu \nu}^+  ,
\ \
H_{\mu \nu}^+ \rightarrow \frac{1}{\theta} \tilde{H}_{\mu \nu}^+ , \ \
\phi \rightarrow \tilde{\phi} \ \  . \label{weight1}
\end{eqnarray}
%where $g \in {\mathbb R}$.
%Putting $g=\theta$,
Note that this changing does not cause nontrivial Jacobian
from the path integral measure because of the BRS symmetry.
Then, the action is rewritten as
\begin{eqnarray}
S \rightarrow \frac{1}{\theta^2} \tilde{S} \ \ \ , \ \ \
L( A_{\mu}, \dots ;
\hat{\partial}_{z_i}, \hat{\partial}_{\bar{z}_i} ) \rightarrow 
\frac{1}{\theta^2} L(\tilde{A}_{\mu} , \dots ; -a_i^{\dagger}
, a_i )\ \ .
\end{eqnarray}
Here the action in the lefthand side
depends on $\theta$ because the derivative is given
by $\partial_{z_i}= - \sqrt{\theta^{ -1}} [ a_i^{\dagger} , \ ] $ and so on.
In contrast, the action $\tilde{S}$ in the righthand side
does not depend on $\theta$ because all $\theta$ parameters are
factorized out.
Using the BRS symmetry or the fact of eq.(\ref{2.13}), it is proved that
the partition function is invariant under the
deformation of $\theta$, because
 $ \delta_{\theta} Z = -2 (\delta \theta ) \theta^{-3}\langle
\tilde{S} \rangle =0$.
$Tr_{{\cal H}_{2dim}} tr (\phi F + \frac{1}{2} \psi \wedge \psi )$
and $tr \phi^n$ are known as observables of Donaldson-Witten theory.
They are rewritten as
$\frac{1}{\theta} Tr_{\cal H} tr ( \tilde{\phi} \tilde{F} +
\frac{1}{2} \tilde{\psi} \wedge \tilde{\psi} )$ and
$tr \tilde{\phi}^n$. 
%%%%%%%% REV5
We use $O$ to represent such observables,
then $\delta_{\theta} \langle O \rangle =0$
are proved in a similar way to the proof of 
$\delta_{\theta} Z=0$.
Therefore, invariance of Donaldson-Witten
theory under $\theta \rightarrow \theta + \delta \theta$ is proved.

We can discuss
the topological twisted ${\mathcal N}=4$ Yang-Mills theory in
noncommutative ${\mathbb R}^4$ similarly
\footnote{There are many kinds of
topological twisted theories of ${\mathcal N}=4$ Yang-Mills theory.
We only consider Vafa-Witten type theory.}
\cite{Vafa-Witten}.
There are additional fields
$(B_{\mu \nu }^+ , c ,H_{\mu})$ and $(\psi_{\mu \nu}^+ , \bar{\eta} , 
\chi_{\mu})$
, where $(B_{\mu \nu }^+ , c ,H_{\mu} )$ are bosonic fields and
$(\psi_{\mu \nu}^+ , \bar{\eta} , \chi_{\mu} )$ are fermionic fields,
where $ B_{\mu \nu }^+ , \psi_{\mu \nu}^+ \in \Omega^{2,+}({\mathbb R}^4 
,\ad P)$.
They are
supersymmetric partners, and
the BRS multiplets are expressed by the following diagram.
\begin{eqnarray}
\begin{array}{ccccc}
&&\psi_{\mu}&& \\
& \hat{\delta}_+ \nearrow &&\searrow \hat{\delta}_-& \\
A_{\mu}&&&&H_{\mu}\\
&\searrow &&\nearrow & \\
&&\chi_{\mu}&&
\end{array}
, \ \ \ \
\begin{array}{ccccc}
&&\psi_{\mu \nu}^+&& \\
& \hat{\delta}_+ \nearrow &&\searrow \hat{\delta}_-& \\
B_{\mu \nu}^+&&&&H_{\mu \nu}^+\\
&\searrow &&\nearrow & \\
&&\chi_{\mu \nu}^+&&
\end{array} \nonumber
\end{eqnarray}
There are two BRS-like operators $\hat{\delta}_+$ and $\hat{\delta}_-$
because of the R-symmetry of the ${\cal N}=4$.
The $\hat{\delta}_+$ transformations are given by
\begin{eqnarray}
\hat{\delta}_+ B_{\mu \nu}^+ = \psi_{\mu \nu}^+ & , &
\psi_{\mu \nu}^+ = i [  B_{\mu \nu}^+ , \phi ]   \\
\hat{\delta}_+ \chi_{\mu} = H_{\mu} & , &
\hat{\delta}_+ H_{\mu} = i [  \chi_{\mu} , \phi ]
\ , \
\hat{\delta}_+ c = \bar{\eta} \ , \
\hat{\delta}_+ \bar{\eta} = i [  c , \phi ] ,
\end{eqnarray}
and the same transformations as (\ref{N=2BRS}) for other fields.
The action of the topological twisted
${\mathcal N}=4$ Yang-Mills theory without the $\tau \int F \wedge F $ is
\begin{eqnarray}
S&=& Tr_{\cal H} tr \ \
 \hat{\delta}_+ \{ \chi^+_{\mu\nu}
\left( H^{+\mu\nu}-i(
F^{+\mu\nu}-i[B^+_{\mu\rho},B^+_{\nu\sigma}] \delta^{\rho\sigma}
%\right. \nonumber \\
%& & \ \ \ \ \ \ \ \ \ \ \ \ \ \ \ \ \ \ \ \ \ \ \ \ \ \ \ \ \ \ \
%\left.
-i[B^+_{\mu\nu},c]) \right)
\}
\nonumber
\\
& &+Tr_{\cal H} tr \ \
\hat{\delta}_+ \{
\chi^{\rho }\left(
H_{\rho}-i(
-2D^\mu B_{\mu\rho}^+
-D_\rho c
)\right)
\}
\nonumber \\
& & +Tr_{\cal H} tr \ \
 \hat{\delta}_+  \{ i[\phi , {\bar \phi}] \eta - i {\bar \eta} [c , {\bar 
\phi}]
+ i [ B^{+ \ \mu \nu} , {\bar \phi}] \psi^+_{\mu \nu} +
 (D_\mu {\bar \phi}) \psi_{\mu}  \} \ .
%&=& (\hbox{Twisted {\cal N}=4 Super Y-M} )
\end{eqnarray}
For this action, we change the variables as
%these fields, we assign the weight of the above transformation as
\begin{eqnarray}
B_{\mu \nu }^+ \rightarrow \frac{1}{\sqrt{\theta}} \tilde{B}_{\mu \nu }^+
 , \ \ \psi_{\mu \nu}^+ \rightarrow \frac{1}{\sqrt{\theta}} 
\tilde{\psi}_{\mu \nu}^+
&, &  c \rightarrow \frac{1}{\sqrt{\theta}} \tilde{c}  , \ \
\bar{\eta} \rightarrow \frac{1}{\sqrt{\theta}}  \tilde{\bar{\eta}}
\ \ \ , \nonumber \\
\chi_{\mu} \rightarrow \frac{1}{\theta} \tilde{\chi}_{\mu}
, \ \
H_{\mu} \rightarrow \frac{1}{\theta} \tilde{H}_{\mu} && \nonumber
\end{eqnarray}
with (\ref{weight1}),
then $S \rightarrow \frac{1}{\theta^2} \tilde{S}$,
and $\tilde{S}$ does not depend on $\theta$.
At last, invariance of the ${\cal N}=4$ topological twisted
theory under $\theta \rightarrow \theta + \delta \theta$ is proved
as same as Donaldson-Witten theory.
%%%%%%%%%%%%%%%%%%%%%%
\\

It is worth commenting on the topological term
$\int F \wedge F$ that exists in usual Vafa-Witten theory
but now is removed.
This term is not written by a BRS exact term,
so we can not adapt above discussion to the topological
term.
But, it is natural that we expect that $\int F \wedge F$
is invariant under the $\theta$ shift.
Indeed, for instanton solutions constructed from
noncommutative deformed ADHM data, we have proof of
invariance of instanton number under $\theta$ shift
\cite{sako3,sako4}.
This is why, we expect that the partition functions
or vacuum expectation values are still invariants even if
the action of the cohomological Yang-Mills theories include
$\int F \wedge F$.
(See also section \ref{modulispace}.)
\\

By applying these facts for several physical models,
some interesting information can be found.
For example, as we will see soon, we can show that
the partition function of the
noncommutative cohomological gauge theory
and the partition function
of the IKKT matrix model
have a correspondence.
This correspondence is not only for certain classical background
theory as we saw in \cite{Aoki-Ishibashi-Iso-Kawai-Tada}.
The reason is as follows.
The IKKT matrix model is
constructed as dimensional reduction of the 10 dimensional
super $U(N)$ Yang-Mills theory with large $N$ limit \cite{IKKT2,IKKT}.
This dimensional reduction is regarded as the large
noncommutative parameter limit ($\theta \rightarrow \infty$ in section \ref{pfn4}).
Taking the large $N$ limit of the matrix model is similar to considering
the Yang-Mills theories on noncommutative Moyal space, i.e.
matrices are regarded as linear operators acting on the Hilbert space
caused from noncommutativity.
By the way, the  noncommutative cohomological Yang-Mills model
on Moyal space in the large $\theta$ limit
is almost the same as the model
of Moore, Nekrasov and Shatashvili (MNS) \cite{Moore-Nekrasov-Shatashvili}.
MNS show that the partition function is calculated
by the cohomological matrix model in \cite{Moore-Nekrasov-Shatashvili}
and related works are seen in
\cite{Bruzzo-Fucito-Morales-Tanzini,Hirano-Kato,Sugino}.
This cohomological matrix model is almost equivalent to the IKKT matrix 
model.
That is why we can produce similar result
by using N.C.cohomological Yang-Mills theories.
To show these facts concretely,
%In this way, there are many interesting subjects to be studied
%by using N.C.cohomological Yang-Mills theory.
we will calculate the partition function of
${\cal N}=4$ d=4 U(1) theory
on N.C.${\mathbb R}^4$ in section \ref{pfn4}
by using the facts given in this section.

%%%%%%%%%%%%%%%%%%%%%%%%%%%%%%%%%%%%%%%%%%%%%%%%%
\section{Universality of Partition Functions}\label{upf}
In this section, we show that the large $\theta$ limit is
equivalent to dimensional reduction.
{}From this fact, we find the universal perspective for the
partition functions of supersymmetric Yang-Mills theories 
in N.C.${\mathbb R}^{2D}$. \\

In the previous section, we consider the case of ${\mathbb R}^4$.
There is two independent noncommutative parameters $\theta^1 , \theta^2$
for the N.C.${\mathbb R}^4$, that is to say, 
after choosing proper coordinate noncommutative parameters are expressed as
\begin{eqnarray}
(\theta^{\mu \nu} )=
\left(
\begin{array}{cc|cc}
0& \theta^1 & 0 & 0 \\
-\theta^1 & 0 & 0& 0 \\
\hline
0 & 0& 0& \theta^2 \\
0& 0& -\theta^2 & 0
\end{array}
\right) \ \ . \label{theta1}
\end{eqnarray}
In the discussion of the previous section,
we take noncommutative parameter shift coincidently,
that is $\theta^1 = \theta^2 =\theta \rightarrow \theta+ \delta \theta $.
However, we can shift $\theta^1 , \theta^2$ independently
without changing partition functions and vacuum expectations.
Further, this discussion is extended to other 
dimensional theories.
\\

Let us consider more general cases than N.C.${\mathbb R}^4$.
Let noncommutative parameter matrix of N.C.${\mathbb R}^{2D}$ be
$ (\theta^{\mu \nu} )= \oplus \theta^i \epsilon^{2i-1 ,\ 2i}$.
In the large $\theta^i$ limit, terms with derivative operators
$\partial_{x_{2i}} :=-i(\theta^i)^{-1} [ x_{2i-1} , * ]$ 
and $-\partial_{x_{2i-1}} :=-i(\theta^i)^{-1} [ x_{2i} , * ]$
%%%%%% REV5
become irrelevant in lagrangians. 
If the partition function and the VEV of arbitrary observables
of the cohomological field theory are well defined,
the terms including 
$\partial_{x_{2i}} $ 
or $\partial_{x_{2i-1}} $
are possible to be removed.
%%%%%%%%%%%%%%%%%%%% REV %%%%%%%%%%%%%%%%%%%%
(In appendix \ref{B}, concrete discussions 
and details are given.)
In the complex coordinate expression, the terms including 
$z_i$ and $\bar{z}_i$ derivatives are omitted. 
Meanwhile, an arbitrary operator is expressed as
\begin{equation}
 \hat{\cal O}=\sum_{n_1,m_1}\cdots\sum_{n_D,m_D}
  {\cal O}_{m_1\cdots m_D}^{n_1\cdots n_D}
  \left|n_1,\cdots,n_D\right>\left<m_1,\cdots,m_D\right|\; ,
\nonumber
\end{equation}
%where we denote $\hat{\cal O}$ 
by using fock space basis.
(See appendix A.)
%operators depend on $n_i$ .  
%%%%%%%%%%%%%%% REV5
We consider a quantum theory of infinite dimensional matrix model, and
${\cal O}_{m_1\cdots m_D}^{n_1\cdots n_D}$ is a variable of path integration. 
Then we cannot distinguish dynamical variables
\begin{eqnarray}
%\sum_{n_1,m_1}\cdots\sum_{n_{i-1},m_{i-1}}
%\sum_{n_{i+1},m_{i+1}}
%\cdots\sum_{n_D,m_D}
  {\cal O}_{m_1\cdots m_{i-1} m_{i+1} \cdots m_D}^{n_1\cdots n_{i-1} n_{i+1} \cdots n_D}
 \left|{\scriptstyle n_1,\cdots,n_{i-1}, n_{i+1}, \cdots , n_D } \right>
  \left< {\scriptstyle m_1,\cdots,m_{i-1}, m_{i+1}, \cdots , m_D }\right| 
\end{eqnarray}
{}from $%\sum_{n_1,m_1}\cdots\sum_{n_D,m_D}
  {\cal O}_{m_1\cdots m_D}^{n_1\cdots n_D}
  \left|n_1,\cdots,n_D\right>\left<m_1,\cdots,m_D\right|$ 
because both of them are infinite dimensional matrices.
%when there is no $\partial_{z_i}$ or $\partial_{\bar{z}_i}$.
{}From the facts 
that there is no $\partial_{z_i}$ or  $\partial_{\bar{z}_i}$ and
it is impossible to distinguish
dynamical variables living in ${\mathbb R}^{2D}$ from variables in 
${\mathbb R}^{2D-2}$,
then we conclude that the large $\theta_i$ limit is
equivalent to the dimensional reduction corresponding to
$x^{2i-1}$ and $x^{2i}$ directions.\\

We have to note two points, here.
The first point is that naive path integrals contain zero mode integrals.
To make story precise, let us define the zero mode here.
Let $\{ \phi_i \}$ be a set of fields and $S[\phi_i]$ be an action functional of
a considered theory. Here, we define the zero mode $\phi_{i}^0$ by $S[\phi_{i}^0]=0$.
To make the partition functions be well defined,
we manage the zero modes, in general.
But it is difficult that the dealing with the zero modes is discussed in general terms.
To avoid this difficulty, the discussion of the zero mode integrals are taken up in the individual cases.
In section \ref{pfn4}, we will closely study the handling of the zero modes
for the case of ${\cal N}=4$ U(1) gauge theory in N.C.${\mathbb R}^4$.

%%%%%%%%%%%%%%%%%%%% REV5 %%%%%%%%%%%
The second point is that 
there might be BPS solutions that become singular at
$\theta^i \rightarrow \infty$ limit.
To the authors' knowledge, such solutions have never been
known until now, but we can not denny their existence.
Since we can not estimate its contribution to the vacuum expectations 
when we calculate them
at the large $\theta^i$ limit, we have to rule out
such singular configurations when we construct the correspondence between
 finite $\theta^i$ and infinite $\theta^i$ .\\

As a summary of these arguments, the following claim is obtained. \\
%%%%%%%%%

%%%%%%%%%%%%%%% REV %%%%%%%%%%%%%%%%
\noindent
{\bf Claim} \\
{\sl
Let $Z_{2D}$ and $\langle O \rangle_{2D}$  be a partition function 
and vacuum expectation value of \ $O$ of a
cohomological field theory in N.C.${\mathbb R}^{2D}$ with $D \ge 1$ such that
$\delta_{\theta} Z_{2D}= 0$ and $\delta_{\theta} \langle O \rangle_{2D} = 0$.
Here, zero mode integrals and contributions from
BPS solutions that become singular at large noncommutative parameter limit
are removed from the path integral of $Z_{2D}$ and $\langle O \rangle_{2D}$.
Let $Z_{2D-2}$ and $\langle O \rangle_{2D-2}$ be the
partition function 
and vacuum expectation value of $O$ of a
noncommutative cohomological field theory in 
N.C. ${\mathbb R}^{2D-2}$, where they are 
given by dimensional reduction of $Z_{2D}$ and $\langle O \rangle_{2D}$.
Then, 
\begin{eqnarray}
Z_{2D} = Z_{2D-2} \; \; \; , \; \; \; 
\langle O \rangle_{2D} = \langle O \rangle_{2D-2}
\ ,
\label{eqclaim}
\end{eqnarray}
i.e. the partition functions of such theories 
do not change
under dimensional reduction from $2D$ to $2D-2$.
}\\
%%%%%%%%%

{}From this claim, we find that
following partition functions of Super Yang-Mills theories on 
N.C. ${\mathbb R}^{2D}$
are equivalent:
\begin{eqnarray}
Z^{8dim}_{{\cal N}=2}
=
Z^{6dim}_{{\cal N}=2}
=
Z^{4dim}_{{\cal N}=4}
=
Z^{2dim}_{{\cal N}=8}
=
Z^{0dim}_{***}
\ , \label{d=8}
\end{eqnarray}
where 
$Z^{Idim}_{
{\cal N}=J}
$ 
is a partition function of the
${\cal N}=J$ super Yang-Mills theory
in noncommutative ${\mathbb R}^I$ with arbitrary gauge group.
They are obtained by dimensional reduction of 
the 8 dimensional ${\cal N}=2$
super Yang-Mills theory. 
Note that the topological terms in the actions of above 
theories should be removed because the topological terms 
is not universal between the different dimensional theories.
The proof of (\ref{d=8}) is as follows.
In the ${\mathbb R}^{2D}$, a topological twist
exists at any time for ${\cal N} \ge 2$.
Using the topological twist, the partition functions are
described as the one of cohomological field theories.
Therefore, $Z^{8dim}_{{\cal N}=2}$ is invariant under $\theta$-shift and 
satisfies the condition of the above claim.
After all, (\ref{d=8}) is obtained. 
We will calculate the partition functions concretely in the case of U(1) in the 
next section.

It is worth adding some comments about above models. 
We consider noncommutative Euclidean spaces.
${\cal N}=4$ super Yang-Mills theory
in N.C. ${\mathbb R}^4$ is given as follows.
At first, we construct the 4-dimensional ${\cal N}=4$ super Yang-Mills theory
by dimensional reduction of the 10 dimensional ${\cal N}=1$
super Yang-Mills defined on Minkowski space
with SO(9,1) symmetry.
In 4-dim, spinor in Euclidean space is defined as well as
the spinor in Minkowski space.
Therefore, we can construct the ${\cal N}=4$ super Yang-Mills theory
in ${\mathbb R}^4$ by formally replacing the metric, gamma matrices and so on.
Since the $\theta$-shift invariance of 
$Z^{4dim}_{{\cal N}=4}$ was shown explicitly in section \ref{ncym}
(see also appendix \ref{B}),
theories connected to the ${\cal N}=4$ $d=4$ super Yang-Mills
theory 
through the dimensional reduction appear in (\ref{d=8}).\\

This discussion is valid not only for the ${\cal N}$=4 case.
For example, we saw that the
$\theta$-shift invariance of $Z^{4dim}_{{\cal N}=2}$
in section \ref{ncym}.
Then, the similar relation should exist :
\begin{eqnarray}
Z^{4dim}_{{\cal N}=2}
= Z^{2dim}_{{\cal N}=4}
= Z^{0dim}_{***}
. 
\end{eqnarray}
\\

Let us summarize this section. 
Universality of 
partition functions and vacuum expectation values of
observables of N.C.cohomological field theories are discussed.
{}From the claim, we found that 
${\cal N} \ge 2$ supersymmetric models
or cohomological field theories in N.C. ${\mathbb R}^{2D}$ 
are invariant under dimensional reduction from $2D$ to $2D-2$.
In the following section, we will apply these facts
to concrete calculations.

%%%%%%%%%%%%%%%%%%%%%%%%%%%%%%%%%%%%%%%%%%%%%%%%%%%%%%%%%%%%
%%%%%%%%%%%%%%%%%%%%%%%%%%%%%%%%%%%%%%%%%%%%%%%%%%%%%%%%%%%%
%%%%%%%%%%%%%%%%%%%%%%%%%%%%%%%%%%%%%%%%%%%%%%%%%%%%%%%%%%%%
\section{${\mathcal N}=4$ $U(1)$ Gauge 
Theory in N.C. ${\mathbb R}^4$}\label{pfn4}

In this section, we calculate the partition function of the
topological twisted ${\cal N}=4$ $U(1)$ gauge theory 
in N.C. ${\mathbb
R}^4$, 
without the
topological term $\int F \wedge F$ in its action.

We perform the calculation in the $\theta \rightarrow \infty$ limit.
The reason why we take this limit is as follows.
As explained in section \ref{ncym},
the partition function and other correlation functions of cohomological
field theories on noncommutative spaces are invariant under the shift
transformation of the noncommutative parameter $\theta$.
So we obtain the exact result by taking $\theta \rightarrow \infty$ limit.
Also this limit makes the calculation executable.

In the operator formalism,
%the topological twisted ${\cal N}=4$ $U(1)$ gauge 
field theories in
N.C. ${\mathbb R}^4$ are expressed as  infinite dimensional matrix models whose
symmetry is $U(N)$ ($N \rightarrow \infty$).
The size of matrices appearing in this model is infinite.
To perform the calculation, we introduce a cut off for the matrix size
. In addition, this matrix model contains trace parts which correspond
to zero modes in $\theta \rightarrow \infty$. Therefore we must carefully treat the trace parts to make
the path integral well-defined.
%\footnote{
%For proper handling of t, see 
%discussions in \cite{Suyama-Tsuchiya}.  
%}.

In subsection \ref{ss_set}, we give the action of the
topological twisted ${\cal N}=4$ $U(1)$ gauge theory in N.C. ${\mathbb R}^4$
in the operator formalism, i.e. in terms of infinite dimensional
matrices.
In subsection \ref{ss_reg}, we truncate the size of the
matrices into finite size, a finite integer $N$.
In subsection \ref{ss_relation},
we explain that the truncated $N \times N$
matrix model action obtained in the previous subsection is equivalent to
 the dimension reduction of the $10$ dim. ${\cal N}=1$ $U(N)$ super
 Yang-Mills action to $0$ dim.
This $U(N)$ matrix model contains traceless parts and trace parts.
In subsection \ref{ss_traceless}, we calculate the partition function
 of the traceless sector. The traceless sector is a $SU(N)$ matrix model.
The partition function of this $SU(N)$ matrix
model was obtained by MNS \cite{Moore-Nekrasov-Shatashvili}.
By modifying their arguments,
we evaluate the $N \rightarrow \infty$ limit of the partition function of
the traceless sector.
In subsection \ref{ss_ex}, we introduce extra parts into the matrices
to handle the trace parts which are zero modes.
The extra parts and trace parts are the next leading terms in the 
$1/\sqrt{\theta}$ expansion.
In \ref{ss_trace}, the calculation of the trace sector is performed.
Our result is presented at the end of this section.

%%%%%%%%%%%
\subsection{Setting}\label{ss_set}
In the Fock space formalism,
i.e. in terms of (infinite dimensional) matrices, the action of the 
topological twisted ${\cal N}=4$ $U(1)$ gauge theory on N.C. ${\mathbb
R}^4$ is expressed as
\begin{eqnarray}
S^{4dim}_{{\cal N}=4} = Tr_{\cal H} \ \ {\hat \delta}_+ &[& + \chi^{+ \ \mu \nu} \{ H^+_{\mu \nu} -i(F^+_{\mu \nu} -i[B^+_{\mu \rho} , B^+_{\nu \sigma}] \delta^{\rho \sigma} -i[B^+_{\mu \nu} , c]) \} \nonumber \\
 & & + \chi^\mu \{ H_\mu - i(-2D^\nu B^+_{\nu \mu} - D_\mu c) \} \nonumber \\
 & & + i[\phi , {\bar \phi}] \eta - i {\bar \eta} [c , {\bar \phi}] + i [ B^{+ \ \mu \nu} , {\bar \phi}] \psi^+_{\mu \nu} +
 (D_\mu {\bar \phi}) \psi_\mu \ \ \ ] .
\label{ncu1-1}
\end{eqnarray}
After acting ${\hat \delta}_+$, (\ref{ncu1-1}) is rewritten as
\begin{eqnarray}
S^{4dim}_{{\cal N}=4} &=&
\nonumber \\
 Tr_{\cal H}  [ && H^{+ \ \mu \nu} \{ H^+_{\mu \nu} -i(F^+_{\mu \nu} -i[B^+_{\mu \rho} , B^+_{\nu \sigma}] \delta^{\rho \sigma} -i[B^+_{\nu \sigma} , c]) \} \nonumber \\
 &+&\chi^{+ \ \mu \nu} \{-i[\chi^+_{\mu \nu} , \phi] +i(2D_\mu \psi_\nu -2i[B^+_{\mu \rho} , \psi^+_{\nu \sigma}] \delta^{\rho \sigma} -i[\psi^+_{\mu \nu} , c] -i[B^+_{\mu \nu} , {\bar \eta}])\} \nonumber \\
 &+&H^\mu \{ H_\mu -i(-2D^\nu B^+_{\nu \mu} - D_\mu c) \} \nonumber \\
 &+&\chi^\mu \{-i[\chi_\mu , \phi] -i(2D^\nu \psi^+_{\nu \mu} +2i[\psi^\nu , B^+_{\nu \mu}] -D_\mu {\bar \eta} +i[\psi_\mu , c]) \} \nonumber \\
 &+&[\phi , {\bar \phi}]^2 +[c , \phi][c , {\bar \phi}] +[B^{+ , \mu \nu} , {\bar \phi}][B^+_{\mu \nu} , \phi] + D^\mu {\bar \phi} D_\mu \phi \nonumber \\
 &+&i[\phi , \eta] \eta +i {\bar \eta} [{\bar \eta} , {\bar \phi}] +i {\bar \eta} [c , \eta] +i[\psi^{+ \ \mu \nu} , {\bar \phi}] \psi^+_{\mu \nu} +i[B^{+ \ \mu \nu} , \eta] \psi^+_{\mu \nu} \nonumber \\
 &+&D^\mu \eta \psi_\mu +i[\psi^\mu , {\bar \phi}] \psi_\mu \ \ \ ] .
\label{ncu1-2}
\end{eqnarray}
{}From (\ref{ncu1-1}) or (\ref{ncu1-2}), we find the BPS equations. For example,
\begin{eqnarray}
 & & F^+_{\mu \nu} -i [B^+_{\mu \rho} , B^+_{\nu \sigma}] \delta^{\rho \sigma} -i [B^+_{\mu \nu} , c] = 0 , \nonumber \\
 & & -2D^\nu B^+_{\nu \mu} -D_\mu c = 0 .
% & & [\phi , {\bar \phi}] = 0 .
\label{bpsncu1}
\end{eqnarray}

In the following, we calculate the partition function
$Z^{4dim}_{{\cal N}=4}$ formally defined as
\begin{equation}
Z^{4dim}_{{\cal N}=4} = \int {\cal D} f e^{-S^{4dim}_{{\cal N}=4}[f]} ,
\label{z0}
\end{equation}
where $f$ means the all matrices $A_\mu , \psi_\mu , \cdots $.
Also we use $f_{boson}$ and $f_{fermion}$ to denotes bosonic
matrices $A_\mu , H_{\mu} ,  \cdots $ and fermionic matrices 
$\psi_\mu , \chi_{\mu} \cdots $, respectively.

In usual commutative spaces, $U(1)$ gauge theories are
free if all matters belong to the adjoint representation, 
because the gauge interactions between the fields belonging to the adjoint
representation are described by commutators of matrices and all
commutators vanish in the $U(1)$ case.
However, in noncommutative spaces, the noncommutativity of the
multiplication induces the $U(1)$ gauge theories to non-Abelian $U(N)$ ($N
\rightarrow \infty$) like gauge theories.   
This $U(N)$ ($N \rightarrow \infty$) is identified with the
unitary transformation group acting on state vectors of the Hilbert space
${\cal H}$
\footnote{It is well known fact that the $U(\infty ) $ is different from
$\lim_{N\rightarrow \infty} U(N)$ , in the meaning of the topology.
In this article, we perform the all calculation by using $\lim_{N\rightarrow \infty} U(N)$, and there is denying that some extra collections appear from the difference.
However, there is no doubt about validity of calculation of $U(N)$ ($N \rightarrow \infty$) as a good approximation even in the case.   }
. 

Let us consider to take the $\theta \rightarrow \infty$ limit
in the calculation of the partition function $Z^{4dim}_{{\cal N}=4}$.
We can evaluate the partition function exactly in this limit,
as explained in section \ref{ncym}. 
% the top of this section.
In the $\theta \rightarrow \infty$ limit we naively expect that all
differential terms in the action vanish and dimensional reduction occur as we
saw in section \ref{upf} .
Therefore, we can perform the calculation by using 
%So the action (\ref{ncu1-1}) or (\ref{ncu1-2}) reduces to
a matrix model in $0$ dim. space whose symmetry is $U(N)$ ($N
\rightarrow \infty$).
We define the action in 0 dim spacetime as
\begin{equation}
S_{MM}^{\infty} = S^{4dim}_{{\cal N}=4} |_{\theta \rightarrow \infty} \ \ : \ \ U(N) \ (N \rightarrow \infty) \ \mbox{matrix model},
\label{smm}
\end{equation}
then, we find $Z^{4dim}_{{\cal N}=4}$ is equal to the
partition function of the matrix model (\ref{smm})
\begin{equation} 
Z^{4dim}_{{\cal N}=4} = \frac{1}{ \mbox{\small $Vol.U(N)(N\!\!\rightarrow \!\!\infty)$}} \int {\cal D} f e^{-S_{MM}^{\infty}} .
\label{zmm}
\end{equation}

To calculate the partition function of this infinite dimensional $U(N)$
($N \rightarrow \infty$) matrix model (\ref{smm}), we need to overcome the
following problems. 

\noindent
(i)The size of the matrices is infinite.
To perform the calculation, we truncate the size of the matrices
into a finite integer $N$.

\noindent
(ii)The matrices contain trace parts. These trace parts play a role of zero modes.
To make the path integral well-defined, we must carefully treat the
trace parts.

In the rest of this section, we solve these problems and obtain the
partition function (\ref{zmm}).

%%%%%%%%%%%
\subsection{Cut off for matrix size}\label{ss_reg}
In this subsection,
we truncate the size of the matrices,
to calculate the partition function.

The Hilbert space of the ${\cal N}=4$ $U(1)$ gauge theory on
N.C.${\mathbb R}^4$ is constructed by a Fock space
\begin{equation}
{\cal H} = \bigoplus_{n_1=0,n_2=0}^{n_1=\infty,n_2=\infty} {\mathbb C}\ |n_1,n_2\rangle .
\label{h}
\end{equation}

We introduce a cut off, a finite integer $N_c$,
and truncate the Hilbert space into a finite dimensional subspace ${\cal
H}_N$ whose dimension is $N$.
We can perform such truncation in several ways.
For example, ${\cal H}_N$ is defined by
\begin{equation}
{\cal H}_N = \bigoplus_{n_1=0,n_2=0}^{n_1={N_c},n_2={N_c}} {\mathbb C} \ |n_1,n_2\rangle .
\label{hn}
\end{equation}
For this case
\begin{equation}
\mbox{dim.} {\cal H_N}= N = (N_c + 1)^2 ,
\label{hnd}
\end{equation}
and the unit matrix of ${\cal H}_N$ is given as
\begin{equation}
{\bf 1}_N = \bigoplus_{n_1=0,n_2=0}^{n_1={N_c},n_2={N_c}} |n_1,n_2 \rangle \langle n_1,n_2|. 
\label{hn1}
\end{equation}
The results and calculations do not
depend on the definition of the cut off in the following discussion.
(See appendix A.)
Therefore we do not use concrete expression of the example
(\ref{hn}).
By definition,
\begin{equation}
Tr_{\cal H} {\bf 1}_N = \mbox{dim.} {\cal H_N}= N .
\label{hn1d}
\end{equation}
For later use, we define ${\cal I}$ as
\begin{equation}
{\cal I} = \frac{1}{\sqrt{N}} {\bf 1}_N ,
\label{calI} 
\end{equation}
which satisfies
\begin{equation}
Tr_{\cal H} \ {\cal I} {\cal I} = 1 .
\label{trII}
\end{equation}

We truncate the infinite dimensional matrices appearing in
(\ref{smm}) into finite dimensional $N \times N$
matrices. We use the symbol $f_N$ to denote the $N \times
N$ truncation of $f$.
For example of (\ref{hn}), if 
$$f=\sum_{n_i=0}^{\infty} \sum_{m_i=0}^{\infty} f^{n_1 n_2}_{m_1 m_2} | n_1 , n_2 \rangle \langle m_1 , m_2 |,  $$
then
$$
f_N=\sum_{n_i=0}^{N_c} \sum_{m_i=0}^{N_c} f^{n_1 n_2}_{m_1 m_2} | n_1 , n_2 \rangle \langle m_1 , m_2 | \ .
$$

Now we consider the finite dimensional $N \times N$ matrix model
$S_{MM}^{N}$ which is obtained by the truncation from (\ref{smm})
\begin{equation}
S_{MM}^{N} = S_{MM}^{\infty} |_{N \times N \ \mbox{truncation}}\  .
\label{smmn}
\end{equation}
%Then let us consider the partition function of the truncated matrix
The partition function of the truncated matrix 
model (\ref{smmn}) is defined by
\begin{equation}
Z^{4dim}_{{\cal N}=4}|_{N} = \frac{1}{Vol.U(N)}  \int {\cal D} f_N e^{-S_{MM}^{N}} .
\label{zmmn}
\end{equation}
%where $f_N$ denotes the $N \times N$ truncation of the infinite
%dimensional matrix $f$.

$N \times N$ matrix $f_N$ is decomposed into the traceless part and the
trace part
\begin{equation}
f_N = f^{su} + f^{tr} ,
\label{fn} 
\end{equation}
where $f^{su}$ is the traceless part and $f^{tr}$ is the trace part.
The traceless part $f^{su}$ is expanded by the generators of the Lie
algebra $su(N)$
\begin{equation}
f^{su} = \sum_{a=1}^{N^2-1} f_{(a)} \tau^a \ , \ \tau^a \in su(N) ,
\label{fsu}
\end{equation}
and $f^{tr}$ is proportional to ${\cal I}$
\begin{equation}
f^{tr} = f_{({\bf 1})} {\cal I} .
\label{ftr}
\end{equation}
The basis, $\tau^a$ and ${\cal I}$, satisfy the following orthonormal 
%orthogonality (normalization) 
conditions
\begin{equation}
Tr_{\cal H} \ \tau^a \ \tau^b = \delta^{a b} \ \ , \ \ Tr_{\cal H} \ {\cal I} \ {\cal I} = 1 \ \ , \ \ Tr_{\cal H} \ \tau^a \ {\cal I} = 0 .
\label{trsu}
\end{equation}

In the naive $\theta \rightarrow \infty$ limit ( i.e. $0$ dimension
reduction ), (\ref{smmn}) contains no trace part $f^{tr}$
\footnote{
Precisely speaking, the trace part of the auxiliary fields appear in
(\ref{smmn}). After integrating out the auxiliary fields, no trace part appears in
(\ref{smmn}).}
\begin{equation}
Z_{MM}^N = Z_{MM}^N |_{{traceless}} \times \int {\cal D} f^{tr} ,
\label{decouple}
\end{equation}
where $Z_{MM}^N |_{{traceless}}$ is defined by
\begin{equation}
Z_{MM}^N |_{{traceless}} = \frac{1}{Vol.SU(N)}
\int {\cal D} f^{su} e^{-S_{MM}^N [f^{su} ]} .
\label{zmmnsu}
\end{equation}
So the trace part $f^{tr}$ plays the role of the zero mode such that
 $S^{N}_{MM}[f^{tr}]=0$.
To make the path integral well-defined, we must carefully treat it. For other handling the zero modes, see for example \cite{Suyama-Tsuchiya}.
However we postpone this task for the moment.
First, we concentrate on the traceless sector.
Before the calculation of the traceless sector,
we explain the equivalence between (\ref{smmn}) and the action considered in
\cite{Moore-Nekrasov-Shatashvili} in the next subsection.

%%%%%%%%%%%%%%%%%%%%%%%%%%%%%%%%%%%%%%%%%%%%%%%%%%%%%%%%%%%%%%%%%%%%%%%%%%%%%%%%%%%%%%%%  4-3
%%%%%%%%%%%
\subsection{Relation to the work of MNS and IKKT }\label{ss_relation}
To explain that the equivalence between (\ref{smmn}) and the action
considered in \cite{Moore-Nekrasov-Shatashvili},
we first recall the fact that the dimensional reduction model from the $D=10$
${\mathcal N}=1$ super Yang-Mills theory to $0$ dimension can 
be reformulated into a
cohomological matrix model \cite{Hirano-Kato,Moore-Nekrasov-Shatashvili}.
%The IKKT matrix model is defined by the $0$ dimension reduction of the $D=10$
%${\mathcal N}=1$ super Yang-Mills theory with $U(N)$ ($N \rightarrow
%\infty$) gauge group.
The $0$ dimension matrix model given 
by dimensional reduction from the $D=10$
${\mathcal N}=1$ super Yang-Mills theory is expressed as
\begin{equation}
S^{10 \rightarrow 0 \ dim}_{{\cal N}=1} = tr \left( \frac{1}{4} [A_M , A_N]^2 + \frac{i}{2} {\bar \Psi} \Gamma^M [A_M , \Psi] \right) ,
\label{sym100}
\end{equation}
where $A_M$ is gauge vector fields and $M,N$ takes $1 \cdots 10$ for the
$10$ dimension Euclid space, or $0, 1 \cdots 9$
for the $10$ dimension Minkowski spacetime.
$\Psi$ is a Majorana-Weyl spinor of the $10$ dimension spacetime. It
contains real 16 components
\footnote{
Note that there is no Majorana-Weyl spinor in 10 dim Euclidean space.
So, if we consider 10 dim model, we should take Minkowski spacetime.
To obtain low dimensional Euclidean model, we first perform dimensional reduction
from 10 dim. to lower dim, and then carry out Wick rotation.  
}
.

%Notice that we can take any gauge symmetry for (\ref{sym100}).
%For example, we can take $U(N)$ or $SU(N)$.
%In the case of $U(N)$, we take the $N \rightarrow \infty$ limit
%to define the IKKT matrix model
%\begin{equation}
%S_{IKKT} = \lim_{N \rightarrow \infty} S^{10 \rightarrow 0 \ dim}_{{\cal N}=1} |_{gauge \ group : U(N)} .
%\label{ikkt}
%\end{equation}

In \cite{Hirano-Kato,Moore-Nekrasov-Shatashvili}, it is shown that
(\ref{sym100}) can be reformulated into a cohomological
matrix model.
The mapping rules between them are as follows
\cite{Moore-Nekrasov-Shatashvili}. $A_M$ are arranged into complex
matrices $\phi$ and $B_i \ (i=1,...,4)$,
\begin{eqnarray}
B_i &=& A_{2i-1} + i A_{2i} \ \ \ (\mbox{for} \ i=1,2,3) , \nonumber \\
B_4 &=& A_9 + i A_{8} , \nonumber \\
\phi &=& A_{7} + i A_{10} ,
\label{mapb}
\end{eqnarray}
and $\Psi$ are arranged as
\begin{equation}
\Psi \rightarrow (\psi_i,\psi^{\dagger}_i) \oplus {\vec \chi} \oplus \eta ,
\label{mapf}
\end{equation}
where ${\vec \chi}$ belongs to the ${\bf 7}$ representation of
$Spin(7)$.
Introducing the bosonic auxiliary matrices ${\vec H}$, we can rewrite
(\ref{sym100}) into a cohomological form
\begin{equation}
S_{MNS} = tr \ {\hat \delta} \left( \frac{1}{16} \eta [\phi , {\bar \phi}] -i {\vec \chi} \cdot {\vec {\cal E}} + {\vec \chi}\cdot {\vec H} + \frac{1}{4} \sum_{a=1}^8 \Psi_a [A_a , {\bar \phi}] \right) ,
\label{mns}
\end{equation}
where ${\vec {\cal E}}$ is defined by
\begin{equation}
{\vec {\cal E}} = \left( [B_i , B_j] + \frac{1}{2} \epsilon_{ijkl} [B^\dagger_k , B^\dagger_l] \ (i < j) \ , \ \sum_i [B_i , B^\dagger_i] \right) .
\label{calE}
\end{equation}
The BRS transformation rules are given as
\begin{eqnarray}
{\hat \delta} A_a = \Psi_a &,& {\hat \delta} \Psi_a = [\phi , A_a] ,
 \nonumber \\
{\hat \delta} {\vec \chi} = {\vec H} &,& {\hat \delta} {\vec H} = [\phi
 , {\vec \chi}] , \nonumber \\
{\hat \delta} {\bar \phi} = \eta &,& {\hat \delta} \eta = [\phi , {\bar
 \phi}] , \nonumber \\
 {\hat \delta} \phi = 0 &.&
\label{brsmns}
\end{eqnarray}
{}From (\ref{mns}) and (\ref{brsmns}), the following BPS equations are obtained
\begin{equation}  
{\vec {\cal E}} = 0 \ \ , \ \  [\phi , {\bar \phi}] = 0 \ \ , \ \ [A_a , \phi]=0 .
\label{bpsmns} 
\end{equation}

One can show that (\ref{mns}) is equivalent to (\ref{smm}),
by using the following correspondence rule \cite{Labastida-Lozano}
\begin{eqnarray}
(\phi \ ,\  c \ , \  \bar{\phi}) &\iff& \left( \sqrt{2} \varphi_{34} \ , \ i \frac{1}{\sqrt{2}} (\varphi_{14}- \varphi_{23})\ , \ \sqrt{2} \varphi_{12} \right) \nonumber \\ 
\label{50}
(B_{\mu \nu}^+ \sigma^{\mu \nu}_{11} \ , \ B_{\mu \nu}^+ \sigma^{\mu \nu}_{12} \ , \ B_{\mu \nu}^+ \sigma^{\mu \nu}_{22} ) &\iff& \left( \sqrt{2} \varphi_{13}\  , \  - \frac{1}{\sqrt{2}} (\varphi_{14}+ \varphi_{23}) \ , \ \varphi_{24} \right),
\end{eqnarray}
where $\varphi$ is defined by
\begin{eqnarray}
\varphi_{k4} = - \varphi_{4k} = \frac{1}{\sqrt{2}} ( A_{k+4} + i A_{k+7} )
\ \ , \ \ 
\varphi_{ij} = ( \epsilon^{ijk} \varphi_{k4} )^* \ \ , \ k=1,2,3 \ .
\label{51}
\end{eqnarray}

Remark that the equivalence among (\ref{smmn}), (\ref{sym100}) and
(\ref{mns}) holds for both $U(N)$ group and $SU(N)$ group.

By choosing gauge group $SU(N)$ and setting $N$ to be a finite integer,
we obtain the equivalence between (\ref{smmn}) and (\ref{mns})
\begin{equation}
S_{MM}^N |_{traceless} = S_{MNS} |_{gauge \ group : SU(N)}^{N:finite} .
\label{seqnsu}
\end{equation}
Therefore
\begin{equation}
Z_{MM}^N |_{traceless} = Z_{MNS} |_{gauge \ group : SU(N)}^{N:finite} ,
\label{zeqnsu}
\end{equation}
where
\begin{eqnarray}
&&Z_{MNS} |_{gauge \ group : SU(N)}^{N:finite} \nonumber \\
&&= \frac{1}{Vol.SU(N)}
\int {\cal D} f^{su} \exp \left\{{-S_{MNS}[f^{su}] |_{gauge \ group : SU(N)}^{N:finite}} \right\} .
\label{zmnssu}
\end{eqnarray}
%Moore {\it et. al.}
MNS obtained the partition function (\ref{zmnssu}) \cite{Moore-Nekrasov-Shatashvili}
\footnote{
See also \cite{Dorey-etc1} where the partition function of the
D-instanton model was calculated.}.

On the other hand,
by choosing gauge group $U(N)$ and taking the $N \rightarrow \infty$
limit, the action
(\ref{sym100}) becomes the IKKT matrix model \cite{IKKT}
\begin{equation}
S_{IKKT} = \lim_{N \rightarrow \infty} S^{10 \rightarrow 0 \ dim}_{{\cal N}=1} |_{gauge \ group : U(N)} .
\label{ikkt}
\end{equation}
So, we obtain the equivalence between (\ref{smm}) and (\ref{ikkt}) ;
\begin{equation}
S_{MM}^{\infty} = S_{IKKT} .
\label{seqinfu}
\end{equation}

%%%%%%%%%%%%%%%%%%%%%%%%%%%%%%%%%%%%%%%%%%%%%%%%%%%%%%%%%%%%%%%%%%%%%%%%%%%%%%%%%%%%%%%%%% 4-4
%%%%%%%%%%%
\subsection{Calculation of traceless sector}\label{ss_traceless}
As explained in the previous subsection
the partition function (\ref{zmmnsu})
is calculated in \cite{Moore-Nekrasov-Shatashvili}.
Their result tells us that
\begin{equation}
Z_{MM}^{N} |_{traceless} = \sum_{d | N} \frac{1}{d^2} ,
\label{zmns}
\end{equation}
where the summation is taken over all divisor $d$ of the finite integer $N$.

One might expect that to obtain the contribution of the traceless part
$f^{su}$ to $Z^{4dim}_{{\cal N}=4}$,
one take the $N \rightarrow \infty$ limit,
\begin{equation}
Z_{MM}^{\infty} |_{traceless} = \lim_{N \rightarrow \infty} Z_{MM}^{N}
|_{traceless} .
\label{zmnslim}
\end{equation}
However $N \rightarrow \infty$ limit in the
righthand side of (\ref{zmns}) is not well-defined.
The reason is as follows.
We see that the righthand side of (\ref{zmns}) is finite; 
\begin{equation}
\sum_{d | N} \frac{1}{d^2} < \sum_{n=1}^{N} \frac{1}{n^2} < 1 +
 \int_1^{\infty} d x \frac{1}{x^2} = 2 .
\label{finite}
\end{equation}
But $\sum_{d | N} \frac{1}{d^2}$ is not a monotonically increasing
function of $N$.
%$\sum_{d | k} \frac{1}{d^2} \le \sum_{d | k'} \frac{1}{d^2}$
%when $k \le k'$ . 
So it does not converge.
For example, if we constrain $N$ to be prime numbers,
\begin{equation}
\lim_{N\rightarrow \infty} \sum_{d | N} \frac{1}{d^2} = 
\lim_{N\rightarrow \infty} ( 1 + N^{-2} )= 1 .
\label{prime}
\end{equation}
If we constrain $N=2^{N'}$,
\begin{equation}
\lim_{N\rightarrow \infty} \sum_{d | N} \frac{1}{d^2} = 
\lim_{N\rightarrow \infty} \sum_{n=0}^{N'} 2^{-2n} = \frac{4}{3} .
\label{2n} 
\end{equation}
Therefore, we must give the proper definition of 
$N \rightarrow \infty$ limit. 

To find a prescription which leads the definite answer of the
$N = \infty$ case,
let us recall the argument of \cite{Moore-Nekrasov-Shatashvili}
where the result (\ref{zmns}) is concluded for a finite $N$.

\noindent
(i)
The authors of \cite{Moore-Nekrasov-Shatashvili} 
separated the coupling constant $g$
to $g$, ${\tilde g}$ and ${\hat g}$
\footnote{In the righthand side of (\ref{mns}), we omitted the coupling constant $g$.},
\begin{equation}
S_{MNS} \rightarrow tr \ {\hat \delta} \left( \frac{1}{16 {\tilde g}} \eta [\phi , {\bar
			\phi}] -i {\vec \chi} \cdot {\vec {\cal E}} + g
			{\vec \chi}\cdot {\vec H} + \frac{1}{4 {\hat g}} \sum_{a=1}^8 \Psi_a [A_a , {\bar \phi}] \right).
\label{mnsr}
\end{equation}

\noindent
(ii)
They deformed the action by redefining ${\cal E}_{ij}$, 
the $({\bf 6} \oplus {\bar {\bf 6}})_{r}$ part of ${\vec {\cal E}}$, as
\begin{eqnarray}
{\cal E}_{ij} &=& \Phi_{ij} - \frac{1}{2} \epsilon_{ijkl}
 \Phi^{\dagger}_{kl}, \nonumber \\
\Phi_{ij} &=& [B_i , B_j] - m \epsilon_{ijk4} B_k,
\label{mdef}
\end{eqnarray}
where $m$ is a mass parameter.
This mass deformation corresponds to the supersymmetry breaking 
from ${\cal N}=4$ to ${\cal N} =1$ in the picture of 4 dimensional
space.

\noindent
(iii)
They again separated the coupling
constants $g$ and ${\hat g}$ as
\begin{eqnarray}
g {\vec \chi}\cdot {\vec H} & \rightarrow & g' \sum_{i < j} \chi_{ij} H_{ij} + g''
 \chi_7 H_7, \nonumber \\
\frac{1}{4 {\hat g}} \sum_{a=1}^8 \Psi_a [A_a , {\bar \phi}] &
 \rightarrow & \frac{1}{4 {\hat g}'} \sum_{a=1}^6 \Psi_a [A_a , {\bar
 \phi}] + \frac{1}{4 {\hat g}''} \sum_{a=7,8} \Psi_a [A_a , {\bar \phi}].
\label{g'g''}
\end{eqnarray}

\noindent
(iv)
Then, they took the following limit,
\begin{equation}
g' \rightarrow 0 \ \ \ \mbox{and} \ \ \ {\hat g}' \rightarrow 0.
\label{gg0}
\end{equation}
Notice that each term in the action is BRS exact. So the partition
function is independent of separated coupling constants $g', g'', \cdots$.
By taking (\ref{gg0}), the partition function is dominated by
configurations around solutions of the following fixed point
equations
\begin{equation}
[ B_i , B_j ] = m \epsilon_{ijk4} B_k \ \ , \ \ [ B_4 , B_i ] = 0 \ \ , \ \ [ B_4 , \phi ] = 0 \ \ \ ,\ \ i=1,2,3\ \ ,
\label{meq}
\end{equation}
where $B_i$, $B_4$ and $\phi$ are all $N \times N$ matrices.

\noindent
(v)
The solution of (\ref{meq}) is given by
\begin{eqnarray}
(B_i)_{N \times N} = (L_i)_{a \times a} \otimes {\bf 1}_{d \times d}, & & \nonumber \\
(B_4)_{N \times N} =
 {\bf 1}_{a \times a} \otimes (B_4)_{d \times d}, 
 & & (\phi)_{N \times N} = {\bf
 1}_{a \times a} \otimes \phi_{d \times d}, 
\label{msol}
\end{eqnarray}
where $a$ is a divisor of $N$ and $d$ is the quotient of $N$ by $a$,
and $(L_i)_{a \times a}$ denotes the generator of the $SU(2)$ group in the $a \times a$ representation.
Of course, there are other solutions of (\ref{meq}),
\begin{eqnarray}
 & & (B_i)_{N \times N} = \left(
\begin{array}{c|c|c}
(L_i)_{a_1 \times a_1} \otimes {\bf 1}_{d_1 \times d_1} & 0 & 0 \\
\hline
 0 & \ddots & 0 \\
\hline
 0 & 0 & (L_i)_{a_k \times a_k} \otimes {\bf 1}_{d_k \times d_k}
\end{array}
\right), \nonumber \\
 & & (B_4)_{N \times N} =
\left(
\begin{array}{c|c|c}
{\bf 1}_{a_1 \times a_1} \otimes (B_4)_{d_1 \times d_1} & 0 & 0
 \\
\hline
0 & \ddots & 0 \\
\hline
0 & 0 & {\bf 1}_{a_k \times a_k} \otimes (B_4)_{d_k \times d_k}
\end{array}
\right), \nonumber \\
 & & (\phi)_{N \times N} = 
\left(
\begin{array}{c|c|c}
{\bf 1}_{a_1 \times a_1} \otimes \phi_{d_1 \times d_1} & 0 & 0 \\
\hline
0 & \ddots & 0 \\
\hline
0 & 0 & {\bf 1}_{a_k \times a_k} \otimes \phi_{d_k \times d_k} 
\end{array}
\right),
\label{msolk}
\end{eqnarray}
where $N = \sum_{l=1}^{k} N_l$, $N_l = a_l \times d_l$.
However as mentioned in \cite{Moore-Nekrasov-Shatashvili} these solutions do
not contribute to the partition function.
The solutions (\ref{msolk}) contain bosonic zero modes, corresponding to
extra $U(1)$ parts $tr_{N_l} \phi, \cdots$, and they are
accompanied by their fermionic partners.
The fermionic partners play a role of fermionic zero modes, and
%fermionic zero modes.
they vanish the path integral.
So the solutions (\ref{msolk}) do not contribute to the partition
function.

\noindent
(vi)
In the above coupling limit (\ref{gg0}) the authors integrated out $B_i$ and
corresponding fermionic partners around the solutions (\ref{msol}) by
the Gaussian integral. The Gaussian integrals from bosons and the one from
fermions cancel each other, so they produce no non-trivial factor.
The resulting {\it effective} action is a matrix model of $d \times d$
matrices, $B_4$, its fermionic partner and $\phi$.

\noindent
(vii)
The partition function of this $d \times d$ matrix model, we call it
$Z_d$, is 
given by 
\begin{equation}
Z_d = \frac{1}{d^2},
\label{zd}
\end{equation}
which is another  
result obtained in the same paper \cite{Moore-Nekrasov-Shatashvili}.
The partition function $Z^{N}_{MM}|_{traceless}$ is given by the sum of
the contributions from the solutions, $Z_d = \frac{1}{d^2}$,
so they concluded (\ref{zmns}).

Now let us turn to the $N \rightarrow \infty$ case. % to define proper $N \rightarrow \infty$ limit.
Our basic strategy is that taking large $N$ limit is done after calculations
with finite $N$. However, the result depends on the definition of the large $N$ limit
as mentioned above. 
To find the proper definition of the large $N$ limit, we consider a naive $N=\infty$ case. That is, 
we do not take the $N \rightarrow \infty$
limit after obtaining the result of the finite $N$ case,
but we take the matrices as $\infty \times \infty$ from the starting point
for a moment.
For the case of $\infty \times \infty$ matrix, the steps (i)-(iv) need no
change, 
but the step (v) should be reconsidered.
In $\infty \times \infty$ matrix, we can embed a solution which has a direct product of
an arbitrary finite dimensional matrix and an infinite dimensional
matrix.
Therefore we obtain solutions,
\begin{eqnarray}
(B_i)_{\infty \times \infty} = (L_i)_{\infty \times \infty} 
\otimes {\bf 1}_{d \times d}, & & \nonumber \\
(B_4)_{\infty \times \infty} =
 {\bf 1}_{\infty \times \infty} \otimes (B_4)_{d \times d}, & & (\phi)_{\infty \times \infty} = {\bf 1}_{\infty \times \infty} \otimes \phi_{d \times d}.
\label{msolinf}
\end{eqnarray}
Now $d$ takes all natural numbers, and $(L_i)_{\infty \times
\infty}$ are the generator of the $SU(2)$ group in the $\infty \times \infty$ representations.
Solutions, like (\ref{msolk}), again do not contribute to the
partition function.
Moreover, one can construct other types of solutions,
\begin{eqnarray}
(B_i)_{\infty \times \infty} = (L_i)_{a \times a} \otimes {\bf
 1}_{\infty \times \infty}, & & \nonumber \\
(B_4)_{\infty \times \infty} =
 {\bf 1}_{a \times a} \otimes (B_4)_{\infty \times \infty}, & & (\phi)_{\infty \times \infty} = {\bf
 1}_{a \times a} \otimes \phi_{\infty \times \infty},
\label{msolinf2}
\end{eqnarray}
and
\begin{eqnarray}
(B_i)_{\infty \times \infty} = (L_i)_{\infty \times \infty} \otimes {\bf
 1}_{\infty \times \infty}, & & \nonumber \\
(B_4)_{\infty \times \infty} =
 {\bf 1}_{\infty \times \infty} \otimes (B_4)_{\infty \times \infty}, & & (\phi)_{\infty \times \infty} = {\bf
 1}_{\infty \times \infty} \otimes \phi_{\infty \times \infty}.
\label{msolinf3}
\end{eqnarray}
%However the contribution from (\ref{msolinf2}) or (\ref{msolinf3}) vanishes, because their contribution $Z_{d = \infty}$ vanishes,
%\begin{equation}
%Z_{d = \infty} = \frac{1}{d^2} \ |_{d = \infty} = 0.
%\label{zdinf}
%\end{equation}
The step (vi), integrating out of $B_i$ and their fermionic partners, again produces no
non-trivial factor, because the cancellation of the Gaussian integrals between bosons and fermions holds for the case of infinite dimensional integral.
Therefore the partition function $Z^{\infty}_{MM}$ is 
given by the sum of contributions from the solutions
(\ref{msolinf},\ref{msolinf2},\ref{msolinf3}),
\begin{equation}
Z^{\infty}_{MM} = Z_{MM}^{(\infty \times d)} + Z_{MM}^{(a \times
 \infty)} + Z_{MM}^{(\infty \times \infty)},
\end{equation}
where the first term in the righthand side comes from (\ref{msolinf}),
the second from (\ref{msolinf2}) and the third from (\ref{msolinf3}).
$Z_{MM}^{(\infty \times d)}$ is still given by the
sum of $Z_d = \frac{1}{d^2}$, as
the step (vii), but in this $N = \infty$ case $d$ runs all natural
numbers ${\mathbb N}$.
%%%%%% REV5
On the other hand, 
it is natural to expect that 
$Z_{MM}^{(a \times \infty)}$ and $Z_{MM}^{(\infty
\times \infty)}$ vanish, because
\begin{equation}
Z_{MM}^{(a \times \infty)} \sim \sum \lim_{d \rightarrow \infty} 
\frac{1}{d^2} = 0\ , \ \ 
Z_{MM}^{(\infty
 \times \infty)} \sim \sum \lim_{d \rightarrow \infty} 
\frac{1}{d^2} = 0,
\end{equation}
if (\ref{zd}) is valid for $d= \infty$.
So we conclude
\begin{equation}
Z^{\infty}_{MM} |_{traceless} = \sum_{d \in {\mathbb N}}
 \frac{1}{d^2} = \zeta (2) = \frac{\pi^2}{6}.
\label{zmnslimr0}
\end{equation}

{}From these considerations, we propose the following definition of
 the large $N$ limit. Let $N(n_i , k)$ be
\begin{eqnarray}
N(n_i , k) \equiv \prod_{i=1}^k  P_i^{n_i}  \ ,
\end{eqnarray}
where $P_i$ are ordered prime numbers, i.e. 
$P_1=2 < P_2 =3 < \cdots < P_k $, and $k$ and $n_i$ are 
positive integers.
We define the large $N$ limit by
\begin{eqnarray}
\lim_{N \rightarrow \infty} \equiv \lim_{k \rightarrow \infty} \lim_{n_i \rightarrow \infty} \ . 
\label{deflargeN}
\end{eqnarray}
Using this definition, we reproduce the same result as (\ref{zmnslimr0}),
\footnote{
It is well known and will be seen in section \ref{modulispace}
that the partition functions of this case
is the sum of the
Euler number of the moduli space, $\chi({\cal M}_k)$ which takes a
rational number in general. So one may expect that $Z_{MM}^{\infty}|_{traceless}$
is given by a rational number. However the summation is an infinite one, then it could take an irrational number, $\frac{\pi^2}{6}$.
}
\begin{eqnarray}
Z_{MM}^{\infty} |_{traceless} = \lim_{k \rightarrow \infty} \lim_{n_i
 \rightarrow \infty} \sum_{l_i = 0}^{n_i} \frac{1}{(\prod_{i=1}^k
 P_i^{l_i})^2} = \prod_{i=1}^\infty \frac{1}{1-P_i^{-2}}= \zeta (2) = \frac{\pi^2}{6}. \nonumber \\
\label{zmnslimr}
\end{eqnarray}

%%%%%%%%%%%%%%%%%%%%%%%%%%%%%%%%%%%%%%%%%%%%%%%%%%%%%%%%%%%%%%%%%%%%%%%%%%%%%%%%%%%%%%%%%%%%%%%% 4-5
%%%%%%%%%%%%%%%%%%%%%%%%
\subsection{Introduction of extra terms}\label{ss_ex}
In this section, we deal with the zero mode problem.
The origin of this problem is the fact that no trace part appears in
(\ref{smmn}).
The reason why all trace parts vanish in (\ref{smmn})
is that we drop all differential terms in the $\theta \rightarrow
\infty$ limit.
To solve the zero mode problem, we keep the next leading terms
including the trace parts in the $1/\sqrt{\infty}$ expansion.

Let us explain the outline of our calculation.
To keep the next leading term, we bring back some extra part $f^{ex}$ living in 
the out side of ${\cal H}_N$. 
%\begin{equation}
%f^{tr} \rightarrow f^{tr} + f^{ex} .
%\label{e}
%\end{equation}
The definition of $f^{ex}$ is given later in this subsection.
Roughly speaking, $f^{ex}$ are matrices appearing in kinetic terms 
$\frac{1}{\theta} f^{ex} \square f^{tr}$ in (\ref{ncu1-2}).
By keeping $f^{ex}$, 
the part of (\ref{ncu1-1}) or (\ref{ncu1-2}) which includes the trace part $f^{tr}$ does not vanish :
\begin{equation}
S_{tr \oplus ex} [f^{tr} , f^{ex}] \ \ = \ \ 
%\lim_{\theta \rightarrow \infty} |_{\mbox{next leading}} \
 S^{4dim}_{{\cal N}=4} |_{\mbox{trace part}}  - O(1/\theta^{1+\epsilon}) \ \ \neq \ \ 0 ,
\label{strex0}
\end{equation}
where $\epsilon$ is an arbitrary positive real number.  
Then the partition function of (\ref{strex0}) is well-defined
\begin{equation}
Z_{tr \oplus ex} \ \ = \ \ \int {\cal D} f^{tr} {\cal D} f^{ex} e^{-S_{tr \oplus ex} [f^{tr} , f^{ex}]} \ \ : \ \ \mbox{well-defined} .
\label{ztrex}
\end{equation}

We suppose $f^{ex}$ has the following expansion form
\begin{equation}
f^{ex} = \sum_{\mu=1}^{4} f_{(\mu)} {\cal T}_\mu .
\label{fex}
\end{equation}
${\cal T}_\mu$ is essentially defined by
the commutator of ${\hat \partial}_\mu$ and ${\bf 1}_N$.
The precise definition of ${\cal T}_\mu$ is as follows.
First of all, we define $T_\mu$ as the commutator of ${\hat \partial}_\mu$ and ${\bf 1}_N$ i.e.
\begin{equation}
T_\mu = [{\hat \partial_\mu} , {\bf 1}_N] .
\label{T}
\end{equation}
In the Fock space formalism,
${\hat \partial}_\mu$ is given as
\begin{eqnarray}
{\hat \partial}_1 = \frac{1}{\sqrt{2\theta^1}}(a_1 - a^\dag_1) &,& {\hat \partial}_2 = \frac{-i}{\sqrt{2\theta^1}}(a_1 + a^\dag_1) , \nonumber \\
{\hat \partial}_3 = \frac{1}{\sqrt{2\theta^2}}(a_2 - a^\dag_2) &,& {\hat \partial}_4 = \frac{-i}{\sqrt{2\theta^2}}(a_2 + a^\dag_2) ,
\end{eqnarray}
where $a_i$ is the annihilation operator and $a^\dag_i$ is the
creation operator.
Given the definition of ${\bf 1}_N$, for example (\ref{hn1}),
we obtain
\begin{eqnarray}
T_1 &=& \frac{\sqrt{N+1} }{\sqrt{2\theta^1}} ( - \sum_{n_2=0}^{N} |N,n_2 \rangle \langle N+1,n_2|  -  \sum_{n_2=0}^{N} |N+1,n_2 \rangle \langle N,n_2|  ) , \nonumber \\
T_2 &=& \frac{-i\sqrt{N+1} }{\sqrt{2\theta^1}} (  -  \sum_{n_2=0}^{N} |N,n_2 \rangle \langle N+1,n_2| 
 +  \sum_{n_2=0}^{N} |N+1,n_2 \rangle \langle N,n_2|  ) , \nonumber \\
T_3 & =& \frac{\sqrt{N+1} }{\sqrt{2\theta^2}} (  -  \sum_{n_1=0}^{N} |n_1,N \rangle \langle n_1,N+1| 
 -  \sum_{n_1=0}^{N} |n_1,N+1 \rangle \langle n_1,N|  ) , \nonumber \\
T_4 &=& \frac{-i\sqrt{N+1} }{\sqrt{2\theta^2}} ( -  \sum_{n_1=0}^{N} |n_1,N \rangle \langle n_1,N+1| 
 + \sum_{n_1=0}^{N} |n_1,N+1 \rangle \langle n_1,N|  ) .
\label{T1234}
\end{eqnarray}
Using (\ref{T1234}), we can show
\begin{equation}
Tr_{\cal H} \ T_\mu T_\nu = \frac{N}{\theta^{i(\mu )}} \delta_{\mu \nu} ,
\label{trexo}
\end{equation}
where $i(\mu )=[ (\mu +1 )/2 ]$ with the symbol 
[ ] indicating a Gaussian symbol. 
${\cal T}_\mu$ is defined by
\begin{equation}
{\cal T}_\mu = \frac{\sqrt{\theta^{i(\mu )}}}{\sqrt{N}} T_\mu ,
\label{calT}
\end{equation}
to satisfy
\begin{equation}
Tr_{\cal H} \ {\cal T}_\mu \ {\cal T}_\nu = \delta_{\mu \nu} .
\label{trex}
\end{equation}

Here we list some formulas about ${\cal I}$ and ${\cal T}_{\mu}$, which will be
used in the calculation of the partition function. They are
\begin{equation}
Tr_{\cal H} \ {\cal I} \ {\cal I} = 1 \ \ , \ \ Tr_{\cal H} \ {\cal T}_\mu \ {\cal T}_\nu = \delta_{\mu \nu} \ \ , \ \ Tr_{\cal H} \ {\cal I} \ {\cal T}_\mu = 0 \ \ \ ,
\label{tr2}
\end{equation}
\begin{eqnarray}
Tr_{\cal H} {\cal I} [{\hat \partial}_\mu , {\cal I}]= 0 &,&
Tr_{\cal H} {\cal I} [{\hat \partial}_\mu , {\cal T}_\nu] = - 
\frac{1}{\sqrt{\theta^{i(\mu )}}} \delta_{\mu \nu} , \nonumber \\
Tr_{\cal H} {\cal T}_\mu [{\hat \partial}_\nu , 
{\cal I}] = + \frac{1}{\sqrt{\theta^{i(\mu )}}} \delta_{\mu \nu} &,& Tr_{\cal H} {\cal T}_\mu [{\hat \partial}_\nu , {\cal T}_\rho ] = 0 \ \ \ ,
\label{tr2d}
\end{eqnarray}
and
\begin{eqnarray}
Tr_{\cal H} {\cal I} [{\cal I} , {\cal I}] = 0 &,& 
Tr_{\cal H} {\cal I} [{\cal I} , {\cal T}_\mu] = 0 , \nonumber \\ 
Tr_{\cal H} {\cal I} [{\cal T}_\mu , {\cal T}_\nu] = 
+ \frac{i \theta^{i (\mu )}}{\sqrt{N}} \theta^{-1}_{\mu \nu} &,& 
Tr_{\cal H} {\cal T}_\mu [{\cal T}_\nu , {\cal T}_\rho] =0 .
\label{tr3}
\end{eqnarray}
For the proof of (\ref{tr2}),(\ref{tr2d}) and (\ref{tr3}), see the
appendix \ref{fock}.
Note that these formulas do not depend on the detail of the definition of the cut off
or (\ref{hn1}).

Remark that, in the $N \rightarrow \infty$ limit, $Tr_{\cal H} {\cal I}
[{\cal T}_\mu , {\cal T}_\nu]$ vanishes,
\begin{equation}
\lim_{N \rightarrow \infty} Tr_{\cal H} {\cal I} [{\cal T}_\mu , {\cal T}_\nu] = 0 .
\label{tr30}
\end{equation}
We will use this $N \rightarrow \infty$ behavior to reduce the
calculation of the partition function to the Gaussian integral. 

%%%%%%%%%%%
\subsection{Calculation of trace and extra sector}\label{ss_trace}
Now, let us calculate the partition function (\ref{ztrex}).
First of all, we list the quantities appearing in the calculation.
Because the model is constructed as a balanced topological field theory,
it is natural to classify them into the BRS multiplets.
For $\{ A_\mu , H_\mu , \psi_\mu , \chi_\mu , H_\mu \}$,
\begin{equation}
\begin{array}{ccccc}
&& \psi_{\mu ({\bf 1})} \ , \ \phi_{\mu (\alpha)} && \\
& {\hat \delta}_+ \nearrow && {\hat \delta}_- \searrow & \\
A_{\mu ({\bf 1})} \ , \ A_{\mu (\alpha)} &&&& H_{\mu ({\bf 1})} \ , \ H_{\mu (\alpha)} \\
& {\hat \delta}_- \searrow && {\hat \delta}_+ \nearrow & \\
&& \chi_{\mu ({\bf 1})} \ , \ \chi_{\mu (\alpha)} && 
\end{array} \ \ \ \ \ , %\alpha = 1,2,3,4 \ ,
\label{multi-A}
\end{equation}
and for $\{ B^+_{\mu \nu} , \psi^+_{\mu \nu} , \chi^+_{\mu \nu} , H^+_{\mu \nu} \}$,
\begin{equation}
\begin{array}{ccccc}
&& \psi^+_{\mu \nu ({\bf 1})} \ , \ \psi^+_{\mu \nu ({\alpha})} && \\
& {\hat \delta}_+ \nearrow && {\hat \delta}_- \searrow & \\
B^+_{\mu \nu ({\bf 1})} \ , \ B^+_{\mu \nu (\mu)} &&&& H^+_{\mu \nu ({\bf 1})} \ , \ H^+_{\mu \nu (\mu)} \\
& {\hat \delta}_- \searrow && {\hat \delta}_+ \nearrow & \\
&& \chi^+_{\mu \nu ({\bf 1})} \ , \ \chi^+_{\mu \nu (\mu)} &&
\end{array} \ \ \ \ \ .
\label{multi-B}
\end{equation}
Note that $A_{\mu ({\bf 1})} $ and $ A_{\mu (\alpha)}$ are coefficients of 
${\cal I}$ and ${\cal T}_{\alpha}$ i.e. 
$A_{\mu} = A_{\mu ({\bf 1})} {\cal I} + \sum_{su(N)} A_{\mu a} \tau^a 
+ \sum A_{\mu (\alpha)} {\cal T}_{\alpha}$,
and other fields are noted by similar manner.

It is necessary to comment on the net components of $\{ A_{\mu (\alpha)}
, \psi_{\mu (\alpha)} , \chi_{\mu (\alpha)} , H_{\mu (\alpha)}\}$ in
(\ref{multi-A}) and $\{ B^+_{\mu \nu (\alpha)} , \psi^+_{\mu \nu
(\alpha)} , \chi^+_{\mu \nu (\alpha)} , H^+_{\mu \nu (\alpha)} \}$ in
(\ref{multi-B}). 
In the following, we use the term $(\mu , \nu ) $ selfdual which means 
that $A_{\mu (\nu)}$ satisfies $A_{\mu (\nu)}= \frac{1}{2}\epsilon_{\mu \nu \rho \sigma} A_{\rho (\sigma)}$.
\\
(i) $\{ A_{\mu (\alpha)} , \cdots \}$ have not sixteen but four components. 
Three of
them satisfy the selfdual relation 
and the rest one is $A_{\mu (\mu)}$ :
\begin{equation}
\{ A_{\mu (\nu)} \ |\ A_{\mu (\nu)}  =  \frac{1}{2} \epsilon_{\mu \nu \rho \sigma} A_{\rho (\sigma)} \  \ (\mu , \nu )\ \mbox{selfdual} \  \}\ 
\mbox{and} \ \{ \sum_{\mu=1}^{4} A_{\mu (\mu)} \} .
\label{netA}
\end{equation}
(ii) $\{ B^+_{\mu \nu (\mu)} , \cdots \}$ have four components
corresponding to $B^+_{\mu \nu (\mu)}$
\begin{equation}
B^+_{\mu \nu (\mu)} = \sum_{\mu=1}^{4} B^+_{\mu \nu (\mu)} .
\label{netB}
\end{equation}

On the other hand, $\{ \phi , c , {\bar \phi} , {\bar \eta} , \eta \}$
contain only trace parts   
\begin{equation}
\begin{array}{ccc}
\phi_{({\bf 1})} && \\
& {\hat \delta}_- \searrow & \\
&& {\bar \eta}_{({\bf 1})} \\
& {\hat \delta}_+ \nearrow & \\
c_{({\bf 1})} && \\
& {\hat \delta}_- \searrow & \\
&& \eta_{({\bf 1})} \\
& {\hat \delta}_+ \nearrow & \\
{\bar \phi}_{({\bf 1})} &&
\end{array} \ \ \ \ \ .
\label{multi-c}
\end{equation}

Later we obtain the Gaussian action (\ref{b1}-\ref{b5},\ref{f1}-\ref{f5},\ref{gf1},\ref{gf2}). For example in (\ref{b1}) we find a term
proportional to 
\begin{equation}
\chi^+_{\mu \nu ({\bf 1})} (A_{\nu (\mu)} - A_{\mu (\nu)}) . \nonumber
\end{equation}
{}From this and other terms in
(\ref{b1}-\ref{b5},\ref{f1}-\ref{f5},\ref{gf1},\ref{gf2}), we see that
the net components (\ref{netA},\ref{netB},\ref{multi-c}) should be taken
to remove the zero modes.

Taking the net components (\ref{netA},\ref{netB},\ref{multi-c}) and
using (\ref{tr2},\ref{tr2d},\ref{tr3}), we obtain 
\begin{eqnarray}
S_{tr \oplus ex} = Tr_{\cal H} \ \ {\hat \delta}_+ &[& + \chi^{+ \ \mu \nu}_{(\rho)} {\cal T}^\rho \{ H^+_{\mu \nu (\sigma)} {\cal T}^\sigma -([{\hat \partial}_\mu , A_{\nu ({\bf 1})} {\cal I}] - [{\hat \partial}_\nu , A_{\mu ({\bf 1})} {\cal I}]) \} \nonumber \\
 & & + \chi^{+ \ \mu \nu}_{({\bf 1})} {\cal I} \{ H^+_{\mu \nu ({\bf 1})} {\cal I} -([{\hat \partial}_\mu , A_{\nu (\rho)} {\cal T}^\rho] - [{\hat \partial}_\nu , A_{\mu (\rho)} {\cal T}^\rho]) \} \nonumber \\
 & & + \chi^\mu_{(\rho)} {\cal T}^\rho \{ H_{\mu (\sigma)} {\cal T}^\sigma - (-2[{\hat \partial}^\nu , B^+_{\nu \mu ({\bf 1})} {\cal I}] - [{\hat \partial}_\mu , c_{({\bf 1})} {\cal I}]) \} \nonumber \\
 & & + \chi^\mu_{({\bf 1})} {\cal I} \{ H_{\mu ({\bf 1})} {\cal I} - (-2[{\hat \partial}^\nu , B^+_{\nu \mu (\rho)} {\cal T}^\rho] - [{\hat \partial}_\mu , c_{(\rho)} {\cal T}^\rho]) \} \nonumber \\
 & & - [{\hat \partial}^\mu , {\bar \phi}_{({\bf 1})} {\cal I}] \psi_{\mu (\nu)} {\cal T}^\nu \ \ \ ] \nonumber \\
 &+& {\cal O}(N^{-\frac{1}{2}}).
\label{strex1}
\end{eqnarray}
Note that $ B^+_{\nu \mu (\rho)}$ looks 12 components but 
only $ B^+_{\nu \mu (\nu)}$ proportional terms survive in
$ Tr_{\cal H}  \chi^\mu_{({\bf 1})} {\cal I} [{\hat \partial}^\nu , B^+_{\nu \mu (\rho)} {\cal T}^\rho] $.
In the $N \rightarrow \infty$ limit,
only quadratic terms survive
\footnote{
Alternatively, we can take the weak coupling limit in the calculation.
In general, partition functions of cohomological field theories are
independent of coupling constants. So they can be evaluated exactly in the
weak coupling limit.}
\begin{equation}
S_{tr \oplus ex}^{\infty} \ = \ \lim_{N \rightarrow \infty} S_{tr \oplus ex} \ \ : \mbox{quadratic action} .
\label{strexlim} 
\end{equation}

The action (\ref{strexlim}) has the following gauge symmetry, 
%\footnote{ 
%If we take anti-selfdual noncommutative parameter $\theta$,
%gauge symmetry can be deformed to 
%$\delta_{gauge} A_{\mu (\nu)} = \frac{1}{\sqrt{\theta_i}} \delta_{\mu \nu}
% \varphi_{(\bf 1)}  +- \frac{\sqrt{\theta_i}}{\sqrt{N}} A_{\mu (\rho)}
% \varphi_{(\bf 1) }\theta^{-1}_{\nu \rho} $. 
%But this difference is ignored in the large $N$ limit.
%}
\begin{equation} 
\delta_{gauge} A_{\mu (\nu)} = \frac{1}{\sqrt{\theta^{i(\mu )}}} \delta_{\mu \nu} \varphi_{(\bf 1)} .
\label{gauge}
\end{equation}
Note that the gauge parameter $\varphi$ contains only one component
$\varphi_{({\bf 1})}$
\begin{equation}
\varphi = \varphi_{({\bf 1})} {\cal I} .
\label{varphi}
\end{equation}

Now we give the BRS transformation rules for $f_{(\bf 1)}$ and
$f_{(\mu)}$.
%Recall the {\it original} BRS transformation rules.
%Except for $A_\mu$ and $\psi_mu$,
%\begin{eqnarray}
%{\hat \delta}_+ {\cal B} = {\cal F} &,& {\hat \delta}_+ {\cal F} = i [{\cal B} , \phi] , \nonumber \\
%{\hat \delta}_+^2 {\cal B} = i [{\cal B} , \phi] &,& {\hat \delta}_+^2 {\cal F} = 0 ,
%\label{brso}
%\end{eqnarray}
%where ${\cal B}$ denotes the bosonic matrix and ${\cal F}$ denotes the
%fermionic one.
%For $A_\mu$ and $\psi_\mu$,
%\begin{eqnarray}
%{\hat \delta}_+ A_\mu = \psi_\mu &,& {\hat \delta}_+ \psi_\mu = [{\hat \partial}_\mu + i A_\mu , \phi] , \nonumber \\
%{\hat \delta}_+^2 A_\mu = [{\hat \partial}_\mu + i A_\mu , \phi] &,& {\hat \delta}_+^2 \psi_\mu = 0.
%\label{brsa}
%\end{eqnarray}
%Using (\ref{tr2}) and (\ref{tr2d}) and the fact that $\phi$ contains the
%trace part $\phi_{({\bf 1})}$ only, we obtain the following
%transformation rules.
Except for $A_{\mu (\nu)} , \psi_{\mu (\nu)}$ and
$A_{\mu ({\bf 1})} , \psi_{\mu ({\bf 1})}$,
\begin{eqnarray}
{\hat \delta}_+ {\cal B}_{(\nu)} = {\cal F}_{(\nu)} &,& {\hat \delta}_+ {\cal F}_{(\nu)} = 0 , \nonumber \\
{\hat \delta}_+^2 {\cal B}_{(\nu)} = 0 &,& {\hat \delta}_+^2 {\cal F}_{(\nu)} = 0 ,
\label{brso1}
\end{eqnarray}
and
\begin{eqnarray}
{\hat \delta}_+ {\cal B}_{({\bf 1})} = {\cal F}_{({\bf 1})} &,& {\hat \delta}_+ {\cal F}_{({\bf 1})} = 0 , \nonumber \\
{\hat \delta}_+^2 {\cal B}_{({\bf 1})} = 0 &,& {\hat \delta}_+^2 {\cal F}_{({\bf 1})} = 0 
\label{brso2}
\end{eqnarray}
where ${\cal B}$ denotes the bosonic matrix and ${\cal F}$ denotes the
fermionic one.

For $A_{\mu (\nu)} , \psi_{\mu (\nu)}$ and
$A_{\mu ({\bf 1})} , \psi_{\mu ({\bf 1})}$,
\begin{eqnarray}
{\hat \delta}_+ A_{\mu (\nu)} = \psi_{\mu (\nu)} &,& 
{\hat \delta}_+ \psi_{\mu (\nu)} = + \frac{1}{\sqrt{\theta^{i(\mu )}}} 
\delta_{\mu \nu} \phi_{\bf 1} , \nonumber \\
{\hat \delta}_+^2 A_{\mu (\nu)} = + \frac{1}{\sqrt{\theta^{i(\mu )}}} 
\delta_{\mu \nu} \phi_{\bf 1} &,& {\hat \delta}_+^2 \psi_{\mu (\nu)} = 0 ,
\label{brsa1}
\end{eqnarray}
and
\begin{eqnarray}
{\hat \delta}_+ A_{\mu ({\bf 1})} = \psi_{\mu \ ({\bf 1})} &,& {\hat \delta}_+ \psi_{\mu ({\bf 1})} = 0 , \nonumber \\
{\hat \delta}_+^2 A_{\mu ({\bf 1})} = 0 &,& {\hat \delta}_+^2 \psi_{\mu ({\bf 1})} = 0 .
\label{brsa2}
\end{eqnarray}

For simplicity, in this section 
we set $\theta^1 = \theta^2 = \theta$ in the following.
Using (\ref{tr2},\ref{tr2d}) and (\ref{brsa1}-\ref{brsa2}),
(\ref{strexlim}) is shown to be
\begin{equation}
S_{tr \oplus ex}^{\infty} = S^{\infty \ boson}_{tr \oplus ex} + S^{\infty \ fermion}_{tr \oplus ex} ,
\label{strex}
\end{equation}
where
\begin{eqnarray}
S^{\infty \ boson}_{tr \oplus ex} &=& + H^{+ \ \mu \nu}_{({\bf 1})} \{ H^+_{\mu \nu ({\bf 1})} + \frac{i}{\sqrt{\theta}}(A_{\nu (\mu)} - A_{\mu (\nu)}) \} \label{b1} \\
 & & + H^{+ \ \mu \nu}_{(\alpha)} \{ H^+_{\mu \nu (\alpha)} - \frac{i}{\sqrt{\theta}}(\delta^\alpha_\mu A_{\nu ({\bf 1})} -\delta^\alpha_\nu A_{\mu ({\bf 1})}) \} \label{b2} \\
 & & + H^\mu_{({\bf 1})} \{ H_{\mu ({\bf 1})} + \frac{i}{\sqrt{\theta}}(-2B^+_{\alpha \mu (\alpha)})\} \label{b3} \\
 & & + H^{\mu}_{(\alpha)} \{ H_{\mu (\alpha)} - \frac{i}{\sqrt{\theta}}(-2B^+_{\alpha \mu ({\bf 1})} + \delta_{\mu \alpha} c_{({\bf 1})}) \} \label{b4} \\
%%%%%%%%%%
 & & + \frac{4}{\theta} {\bar \phi}_{({\bf 1})} \phi_{({\bf 1})} \label{b5} ,
\end{eqnarray}
and
\begin{eqnarray}
S^{\infty \ fermion}_{tr \oplus ex} &=& - \frac{i}{\sqrt{\theta}} \chi^{+ \ \mu \nu}_{({\bf 1})} (\psi_{\nu (\mu)} - \psi_{\mu (\nu)}) \label{f1} \\
 & & - \frac{2i}{\sqrt{\theta}} \chi^{+ \ \mu \alpha}_{(\alpha)} \psi_{\mu ({\bf 1})} \label{f2} \\
 & & + \frac{i}{\sqrt{\theta}} \chi^\mu_{({\bf 1})} (2\psi^+_{\alpha \mu (\alpha)}) \label{f3} \\
 & & - \frac{i}{\sqrt{\theta}} \chi^\mu_{(\alpha)} (2\psi^+_{\alpha \mu ({\bf 1})} + \delta_{\mu \alpha} {\bar \eta}_{({\bf 1})}) \label{f4} \\
%%%%%%%%%%
 & & + \frac{i}{\sqrt{\theta}} \eta_{({\bf 1})} \psi^{\mu}_{(\mu)} \label{f5} .
\end{eqnarray}

Now we fix the gauge symmetry (\ref{gauge}).
We introduce the ghost $\rho$, the anti-ghost ${\bar \rho}$ and the
Nakanishi-Lautrup field $b$.
Their ghost number are assigned as
($+1$,$-1$,$0$) for ($\rho$,${\bar \rho}$,$b$), respectively.
BRS transformations for $\{ {\bar \rho} , b , \rho
\}$
are defined as
\begin{equation}
{\hat \delta}_+ b = \rho \ , \  {\hat \delta}_+ \rho = 0 \ , \ 
{\hat \delta}_+ \bar{\rho} =0 .
%\begin{array}{ccc}
%&& {\bar \rho}_{({\bf 1})} \\
%& {\hat \delta}_+ \nearrow & \\
%b_{({\bf 1})} && \\
%& {\hat \delta}_- \searrow & \\
%&& \rho_{({\bf 1})}
%\end{array} \ \ \ \ \ .
\label{multi-b}
\end{equation}
Because the gauge symmetry is given by (\ref{gauge}),
$\{ {\bar \rho} , b , \rho \}$ contain only the trace parts.
%$ \delta^{[\rho]}_{gauge}$ denotes the gauge transformation with
%the transformation parameter $\rho$.

Let us introduce a gauge fixing action by
\begin{equation}
S_{g.f.} = Tr_{\cal H} \ \ {\hat \delta}_+ \ \ [ \ \ {\bar \rho}_{({\bf 1})} {\cal I}
(b_{({\bf 1})} {\cal I} + [{\hat \partial}^\mu , A_{\mu (\nu)} {\cal T}^\nu]) \ \ \ ] .
\label{strexgf0}
\end{equation}
To get the BRS exact action including the gauge fixing action, 
let us deform
the BRS transformation rules for $A_{\mu
(\nu)},\psi_{\mu (\nu)}$ (\ref{brsa1}) as
%are deformed to
%\begin{eqnarray}
%{\hat \delta}_+ A_\mu = \psi_\mu + \delta^{[\rho]}_{gauge} A_\mu &,&
%{\hat \delta}_+ \psi_\mu = D_\mu \phi + \delta^{[\rho]}_{gauge} \psi_\mu , \nonumber \\
%{\hat \delta}_+ {\bar \rho} = b &,& {\hat \delta}_+ b = 0 , \nonumber \\
%{\hat \delta}_+ \rho = 0 &.&
%\label{brsg}
%\end{eqnarray}
%\begin{eqnarray}
%{\hat \delta}_+ A_{\mu ({\bf 1})} = \psi_{\mu ({\bf 1})} & , & {\hat \delta}_+ \psi_{\mu ({\bf 1})} = 0 , \nonumber \\
%{\hat \delta}_+ b_{({\bf 1})} = {\bar \rho}_{({\bf 1})} & , & {\hat \delta}_+ {\bar \rho}_{({\bf 1})} = 0 , \nonumber \\
%{\hat \delta}_+ \rho_{({\bf 1})} = 0 & , &
%\label{brsgr1}
%\end{eqnarray}
%and
\begin{eqnarray}
{\hat \delta}_+ A_{\mu (\nu)} &=& \psi_{\mu (\nu)} + \frac{1}{\sqrt{\theta}} \delta_{\mu \nu} \rho_{({\bf 1})}  \nonumber \\
%- \frac{\sqrt{\theta_i}}{\sqrt{N}} A_{\mu (\rho)} \rho_{({\bf 1})} \theta^{-1}_{\nu \rho} , \nonumber \\
{\hat \delta}_+ \psi_{\mu (\nu)} &=& + \frac{1}{\sqrt{\theta}} \delta_{\mu \nu} \phi_{({\bf 1})} .
% - \frac{\sqrt{\theta_i}}{\sqrt{N}} \psi_{\mu (\rho)} \rho_{({\bf 1})} \theta^{-1}_{\nu \rho} .
\label{brsa1r}
\end{eqnarray}
%In the $N \rightarrow \infty$ limit, (\ref{brsgr2}) becomes
%\begin{eqnarray}
%{\hat \delta}_+ A_{\mu (\nu)} &\longrightarrow& \psi_{\mu (\nu)} +
% \frac{1}{\sqrt{\theta_i}} \delta_{\mu \nu} \rho_{({\bf 1})} , \nonumber \\
%{\hat \delta}_+ \psi_{\mu (\nu)} &\longrightarrow& + \frac{1}{\sqrt{\theta_i}}
% \delta_{\mu \nu} \phi_{({\bf 1})} .
%\label{brsa1r0}
%\end{eqnarray}
%then
%\begin{eqnarray}
%{\hat \delta}_+ A_{\mu (\mu)} &\longrightarrow& - \frac{4i}{\sqrt{\theta_i}} \rho_{({\bf 1})} , \nonumber \\
%{\hat \delta}_+ \psi_{\mu (\mu)} &\longrightarrow& - \frac{4i}{\sqrt{\theta_i}} \phi_{({\bf 1})} .
%\label{brsgr204}
%\end{eqnarray}
(\ref{strexgf0}) is rewritten into
\begin{eqnarray}
S_{g.f.} &=& + b_{({\bf 1})} (b_{({\bf 1})} - \frac{1}{\sqrt{\theta}} A_{\mu , (\mu)}) \label{gf1} \\
 & & + \frac{4}{\theta} {\bar \rho}_{({\bf 1})} \rho_{({\bf 1})} + \frac{1}{\sqrt{\theta}} {\bar \rho}_{({\bf 1})} \psi_{\mu (\mu)} \label{gf2} .
\end{eqnarray}

%Because (\ref{b1}-\ref{b5}),(\ref{f1}-\ref{f5}) and
%(\ref{gf1},\ref{gf2}) are all quadratic, 
%we can obtain the partition function by performing Gauss integral.
We list degrees of the Gaussian integral in
(\ref{b1}-\ref{b5}),(\ref{f1}-\ref{f5}) and (\ref{gf1},\ref{gf2}).

\begin{equation}
\begin{array}{ccc}
 & \mbox{from bosons} & \\
\mbox{degree} & & \\
3+3 & H^+_{\mu \nu ({\bf 1})} \ \ , \ \ A_{\mu (\nu)} \ 
\ (\mu , \nu )\ \mbox{selfdual} \   & \mbox{in} \ (\ref{b1}) \\
4+4 & H^{+ \ \mu \nu}_{(\nu)} \ \ , \ \ A_{\mu ({\bf 1})} & \mbox{in} \ (\ref{b2}) \\
4+4 & H^\mu_{({\bf 1})} \ \ , \ \ B^{+ \ \alpha \mu}_{(\alpha)} & \mbox{in} \ (\ref{b3}) \\
3+1+3+1 & H_{\mu (\alpha)} \ 
\ (\mu , \alpha )\ \mbox{selfdual} \  \ 
, \ \ H_{\mu (\mu)} \ \ , \ \ B^+_{\alpha \mu ({\bf 1})} \ \ , \ \ c_{({\bf 1})} & \mbox{in} \ (\ref{b4}) \\
1+1 & \phi_{({\bf 1})} \ \ , \ \ {\bar \phi}_{({\bf 1})} & \mbox{in} \ (\ref{b5}) \\
1+1 & b_{({\bf 1})} \ \ , \ \ A_{\mu (\mu)} & \mbox{in} \ (\ref{gf1}) 
\end{array} 
\label{degb}
\end{equation}

\begin{equation}
\begin{array}{ccc}
 & \mbox{from fermions} & \\
\mbox{degree} & & \\
3+3 & \chi^+_{\mu \nu ({\bf 1})} \ \ , \ \ \psi_{\nu (\mu)} \ 
\ (\mu , \nu )\ \mbox{selfdual} \  
 & \mbox{in} \ (\ref{f1}) \\
4+4 & \chi^{+ \ \mu \alpha}_{(\alpha)} \ \ , \ \ \psi_{\mu ({\bf 1})} & \mbox{in} \ (\ref{f2}) \\
4+4 & \chi^{\mu}_{({\bf 1})} \ \ , \ \ \psi^{+ \ \alpha \mu}_{(\alpha)} & \mbox{in} \ (\ref{f3}) \\
3+1+3+1 & \chi_{\mu (\nu)} \ 
\ (\mu , \nu )\ \mbox{selfdual} \  
 \ , \ \ \chi^\mu_{(\mu)} \ \ , \ \ \psi^+_{\alpha \mu ({\bf 1})} \ \ , \ \ {\bar \eta}_{({\bf 1})} & \mbox{in} \ (\ref{f4}) \\
1+1 & \eta_{({\bf 1})} \ \ , \ \ \psi^\mu_{(\mu)} & \mbox{in} \ (\ref{f5}) \\
1+1 & \rho_{({\bf 1})} \ \ , \ \ {\bar \rho}_{({\bf 1})} & \mbox{in} \ (\ref{gf2})
\end{array}
\label{degf}
\end{equation}

{}From (\ref{degb}) and (\ref{degf}), we see that the path integral
contains no zero mode, so we obtain a definite partition function.
%%%%%%%%%%%%%%%%%%%%%%%%%%
%%% @@@@@@@ syuusei 1
%%%%%%%%%%%%%%%%%%%%%%%%%%
We adopt a standard path integral measure, which is largely
expressed by
\begin{equation}
{\cal D} f = \prod \frac{d f_{boson}}{\sqrt{2 \pi}} \prod d f_{fermion} ,
\label{measure}
\end{equation}
where $f_{boson}$ denotes a bosonic field and $f_{fermion}$ denotes a
fermionic field.
For the precise definition of ${\cal D} f$ and the validity of this
choice, see the next subsection and appendix \ref{norm}.
Then $Z_{tr \oplus ex}^{\infty}$ is calculated as $1$,
\begin{equation} 
Z_{tr \oplus ex}^{\infty} = \int {\cal D} f e^{ - ( S_{tr \oplus ex}^{\infty} + S_{g.f.} )} = 1 .
\label{ztrexgf=1}
\end{equation}

%%%%%%%%%%%%%%%%%%%%%%%%%%
%%% @@@@@@@@ syuusei 2
%%%%%%%%%%%%%%%%%%%%%%%%%%
\subsection{Results and remarks of this section} \label{randr}
{}From (\ref{zmnslimr}) and (\ref{ztrexgf=1}), we conclude that
the partition function of the ${\mathcal N}=4$ $U(1)$ gauge
theory on N.C. ${\mathbb R}^4$ is given by 
\begin{equation}
Z^{4dim}_{{\mathcal N}=4} = Z_{MM}^{\infty} = Z_{MM}^{\infty} |_{{traceless}}  \times Z_{tr \oplus ex}^{\infty} =  \frac{\pi^2}{6} .
\label{zn4}
\end{equation}

We comment on the universality of partition
function, (\ref{d=8}).
Our calculation consists of largely two steps.
In the first step the traceless part is treated,
then in the second step
the trace and extra parts are managed.
The first step is manifestly dimensionally independent, because
after the dimensional reduction to 0 dim 
all actions of (8-dim , ${\mathcal N}=2$),
(6-dim , ${\mathcal N}=2$),(4-dim , ${\mathcal N}=4$) 
and (2-dim , ${\mathcal N}=8$) are the same as the IKKT matrix
model action.
On the other hand, the calculations in the second step may seem to
depend on the dimension of the model, since we keep the derivatives,
${\hat \partial}_\mu$. 
However, the same result $Z_{tr \oplus ex}^{\infty} = 1$ is expected to
be universal.
The reason is as follows.
The second step, introducing the extra part and
fixing the gauge symmetry (\ref{gauge}), is a kind of regularization of the zero mode integral.
As expected from other regularization method, for example, naively
dropping the trace part, equivalent to dividing the path integral
measure by the $U(1)$ gauge volume,
the regularization should produce a trivial factor $1$.
In our regularization method, this is implemented by the supersymmetry.
Also, as explained in appendix \ref{norm}, our regularization
is valid for all of
(8-dim , ${\mathcal N}=2$), (6-dim , ${\mathcal N}=2$),
(4-dim , ${\mathcal N}=4$) and (2-dim , ${\mathcal N}=8$).
Then we conclude
\begin{equation}
Z^{8dim}_{{\mathcal N}=2} =Z^{6dim}_{{\mathcal N}=2} =Z^{4dim}_{{\mathcal N}=4} =Z^{2dim}_{{\mathcal N}=8} = \frac{\pi^2}{6}.
\label{d=8r}
\end{equation}

Finally, we make a remark relating the mathematical
significance of (\ref{d=8r}).
In topological field theories, the path integral can be decomposed into finite
dimensional integrals of the moduli space defied by the BPS equations and infinite dimensional
integrals of fluctuations around each vacuum.
The absolute value of the infinite dimensional integrals of the fluctuations should be normalized to $1$ to make the partition functions welldefined,
then only the integrals of the moduli space remain. 
(See also appendix \ref{norm}.)
If the moduli space is compact,
the remained moduli integrals produce a definite number,
which is the Euler number of some vector bundle over the
moduli space.
In the case of this section, each $1/d^2$ in (\ref{zmnslimr0})
corresponds to the Euler number.
In this light,
our prescription above, adopting the measure
(\ref{measure}) to obtain (\ref{ztrexgf=1}), 
is an almost unique choice,
though it may seem to be chosen by hand.
Also, for the traceless part, the similar prescription is
performed in \cite{Moore-Nekrasov-Shatashvili}.
To conclude, the result $\pi^2 / 6$ is decided without ambiguity and has an absolute meaning as a topological invariant.

%%%%%%%%%%%%%%%%%%%%%%%%%%%%%%%%%%%%%%%%%%%%%%%%%%%%%%%%%%%%%%
%%%%%%%%%%%%%%%%%%%%%%%%%%%%%%%%%%%%%%%%%%%%%%%%%%%%%%%%%%%%%%
%%%%%%%%%%%%%%%%%%%%%%%%%%%%%%%%%%%%%%%%%%%%%%%%%%%%%%%%%%%%%%

%%%%%%%%
%%%%%%%%
\section{Moduli Space and Instanton Number}\label{modulispace}
In this section, we concentrate on the relation 
between the moduli space
of the Monads and the partition function of the ${\cal N}=4$
supersymmetric Yang-Mills theory.
The partition function of Vafa-Witten theory
is given by the generating function of the Euler number of the 
some vector bundle over the moduli space with sign $\pm 1$:
\begin{eqnarray}
Z&=&\sum_{k=1} \epsilon_k \ \chi({\cal M}_k) q^k \\
q^k&=&e^{2 \pi i k \tau} \ , \ \ \epsilon_k = \pm 1.
\end{eqnarray}
Here $\tau$ is the complex coupling constant
and ${\cal M}_k$ is the moduli space defined by
\begin{eqnarray}\label{moduli1}
\left\{ A , B , c  | 
F^{+\mu\nu}-i[B^+_{\mu\rho},{B^+_{\nu}}^{\rho}] 
-i[B^+_{\mu\nu},c]=0 \; , 
 2D^\mu B_{+\mu\rho}
+D_\rho c =0 \right\}/{\cal G},
\end{eqnarray}
where ${\cal G}$ is the gauge transformation group.
In addition, if $\chi^{\mu \nu} , \chi^{\mu}$  zero-modes are sections of the 
cotangent bundle of ${\cal M}_k$, then 
$\chi({\cal M}_k)$ is the Euler number of ${\cal M}_k$.
Particularly, the base 4-fold satisfies the vanishing theorem
in \cite{Vafa-Witten}, then the moduli space is
identified with the instanton moduli space with its instanton
number $k$.
Therefore, it is important to investigate the ${\cal M}_k$.
\\

It is natural to assume that the topology of the moduli space 
does not change under the $\theta$-shift.
After dimensional reduction (large $\theta$ limit) ,
let us replace variables as (\ref{mapb}), 
(\ref{50}) 
and (\ref{51}).
%we introduce following variables:
%\begin{eqnarray}
%A_{2i-1}+ i A_{2i} = B_i \; , \; i=1\sim 4 \ .
%\end{eqnarray}
%Here we use 10 dimensional gauge theory representation.
%The correspondence between $A_k , k=5, \cdots , 10$
%and topological twisted ${\cal N}=4$ 4-dim variables is as follows
%\cite{Labastida-Lozano}.
%\begin{eqnarray}
%(\phi \ ,\  c \ , \  \bar{\phi}) &\iff& 
%\left( \varphi_{12} \ , \  i \frac{1}{2}(\varphi_{14}- \varphi_{23})\ ,\  
%\varphi_{34} \right) \\
%(B_{\mu \nu}^+ \sigma^{\mu \nu}_{11} \ , \  
%B_{\mu \nu}^+ \sigma^{\mu \nu}_{12} \ , \ 
% B_{\mu \nu}^+ \sigma^{\mu \nu}_{22} ) &\iff&
% \left( \varphi_{34}\  , \  \frac{1}{2}(\varphi_{14}+ \varphi_{23})
% \ , \  \varphi_{24}\right),
%\nonumber 
%\end{eqnarray}
%where $\varphi$ is defined by
%\begin{eqnarray}
%\varphi_{k4}=-\varphi_{4k}:= \frac{1}{\sqrt{2}}( A_{k+4} + i A_{k+7} )
%\ \ , \ \ 
%\varphi_{ij} := (\epsilon^{ijk} \varphi_{k4})\ , \ k=1,2,3 \ . \\
%\end{eqnarray}
As operators, fields are infinite dimensional matrices.
If matrix size of these $B_i$ is cut off at $N$ ,
BPS eqs. (\ref{bpsncu1}) are replaced by hyperk\"ahler momentum maps
\begin{eqnarray}\label{monad}
\mu_{\mathbb C}:=[ B_i , B_j ] + \frac{1}{2}
\epsilon_{ijkl}[ B_k^{\dagger}, B_l^{\dagger} ] =0, \nonumber \\
\mu_{\mathbb R}:=\sum_{i=1}^4 [B_i , B_i^{\dagger} ] =0 ,
\end{eqnarray}
then the moduli space is determined by
\begin{eqnarray}\label{moduli2}
{\cal M}_N= ( \mu_{\mathbb C}^{-1}(0) \cap \mu_{\mathbb R}^{-1}(0) )/ U(N) .
\end{eqnarray}
It is known that the solutions of eqs.(\ref{monad}) 
include the solutions of simultaneous ADHM eqs. \cite{Park}.
$\theta$ deformation realizes the continuous connection between
(\ref{moduli1}) and (\ref{moduli2}).
This is a direct correspondence between 
BPS equations of
noncommutative field theory and Monads 
by means of changing the noncommutative parameter.
\\

Turning now to next issue, let us study the partition function whose action
functional includes the topological term. 
In section \ref{pfn4}, we perform the calculation with
the action functional which does not include the term 
of 
$ \tau \int F \wedge F$ ( or $ \tau \ Tr_{\cal H} F \wedge F$ ).
%{}From the facts of \cite{Park}, 
%we expect that $ \int F \wedge F = 16 \pi^2 n$
%for the solution corresponding ${\cal M}_n$. 
In the MNS calculation,
they use the mass deformation to decompose the theory
to more simple ones whose partition function is
given by $1/d^2$ in (\ref{zmns}).
(See section 7 in \cite{Moore-Nekrasov-Shatashvili} and section 
\ref{ss_traceless} in this article. )
This mass deformation causes supersymmetry breaking from
${\cal N}=4$ to ${\cal N}=1$.
$B_1 , B_2 , B_3$ become massive, and $B_4 , B_4^{\dagger}$ 
and $\phi , \bar{\phi}$ are left for massless fields.
If we consider this mass deformation 
in the finite $\theta$ theory, we find that gauge fields are given from
$B_4 , B_4^{\dagger}$ 
and $\phi , \bar{\phi}$ as 4-dim theory,
because the massless fields correspond 
to the unbroken gauge fields.
In the reduced theory after integrating out $B_1, B_2 $ and $B_3$,
fixed point loci are defined by
\begin{eqnarray}
 [ B_4 , B_4^{\dagger} ] =0 \ , \ [ \phi , \bar{\phi} ] =0
\ , \ [ B_4 , \phi ] = 0, \label{d=4MNS}
\end{eqnarray}
where $B_4 , B_4^{\dagger}$ 
and $\phi , \bar{\phi}$ are $d \times d $ matrices where
$d$ is a divisor of $N$ and is appearing in the argument of (\ref{zmns}).
Furthermore, contributions for the partition function
are given by isolated fixed points, as MNS mentioned in the end of section 5
in \cite{Moore-Nekrasov-Shatashvili}. 
At least one of $B_4$ and $\phi$ is the rank $d$ , 
when $B_4$ and $\phi$ are solutions of the fixed point equations and the fixed points contribute to the path integral.
Because if $\mbox{rank} < d$ then there are zero modes 
of the equations
\begin{eqnarray}
 [ \delta B_4 , B_4^{\dagger} ] +  [ B_4 , \delta  B_4^{\dagger} ] =0 \ , 
\ [ \delta \phi , \bar{\phi} ] +[ \phi , \delta \bar{\phi} ] =0
\ , \ [ \delta B_4 , \phi ]+ [ B_4 , \delta \phi ]  = 0, \label{d=4MNS2}
\end{eqnarray}
where these equations are given by variation of (\ref{d=4MNS}).
These zero modes mean that the fixed point loci are non-zero dimension
and path integrals vanish by the fermionic zero modes.
With attention to these points, 
if we specify the instanton numbers corresponding to solutions of (\ref{d=4MNS})
labeled by $d$, then we determine the partition function whose action
functional includes the topological term.\\

A hint to speculate the instanton number is ADHM correspondence.
The solutions of (\ref{d=4MNS}) is included 
in the set of solutions of noncommutative deformed 
ADHM eqs. corresponding to $d$ instanton, i.e.
\begin{eqnarray}\label{ADHM} 
[ B_4 , B_4^{\dagger} ] +[ \phi , \bar{\phi} ] + II^{\dagger}-J^{\dagger}J
=0 \ , \ 
[ B_4 , \phi ] + IJ= 0 ,
\end{eqnarray}
where $I$ and $J^{\dagger}$ are $d$-dim. vectors.
This is ADHM equations of noncommutative U(1) theories
under the condition of noncommutativity $\theta^1 = -\theta^2$
\cite{Nekrasov-Schwarz}.
Here we have to fix $I$ and $J^{\dagger}$ to
compare (\ref{ADHM}) with (\ref{d=4MNS}) as
\begin{eqnarray}
I= 0_d \ \ ,\ \ J^{\dagger}=0_d ,
\end{eqnarray}
where $0_d$ is 0 vector of $d$-dim.
Then, the solutions of (\ref{ADHM}) are given by 
the solutions of (\ref{d=4MNS}).
{}From this observation, we find that the moduli space of 
$B_4 , B_4^{\dagger}$ 
and $\phi , \bar{\phi}$ , which are gauge fields in this case, 
is the submanifold in instanton 
moduli space of instanton number $d$.

%The other reason is given from the D-instanton view point \cite{Dorey-etc2}.
%Note that the there is no direct 
%correspondence of \cite{Dorey-etc1}'s 
%ADHM data and $B_4$ and $\phi$, so it is not easy to compare
%the partition function in \cite{Dorey-etc1} with our partition function.
%However it seems reasonable to suppose that 
%the diagonal elements are interpreted as the position of the
%$d$ instantons and the instanton number
%is $d$ in this case. 

Therefore, someone might think it is not so strange to expect that
the instanton number is given as
$
-\frac{det(\theta)^{\frac{1}{2}}}{16 \pi^2 } Tr_{\cal H} F \wedge F = d \ , 
$
where the gauge fields correspond to 
$B_4 , B_4^{\dagger}$ 
and $\phi , \bar{\phi}$, and
we conjecture that 
the partition function of the ${\cal N}=4$ 
U(1) gauge theory in noncommutative ${\mathbb R}^4$
with the topological term $ \tau \int F \wedge F$ is given by
$
\tilde{Z}_{{\cal N}=4, \tau}^{4dim} = 
 \sum_{d=0}^{\infty} \frac{1}{d^2} 
e^{2\pi i \tau d} \ . 
$
However, 
It would still be unwise to conclude $
\tilde{Z}_{{\cal N}=4, \tau}^{4dim} = 
 \sum_{d=0}^{\infty} \frac{1}{d^2} 
e^{2\pi i \tau d} \ 
$
, because
the direct corresponding with the instanton number
and $B_4 , B_4^{\dagger}$ 
and $\phi , \bar{\phi}$ fixed point locus labeled by $d$
is unknown.
%We leave constructing the direct correspondence for the future work.
Meanwhile,
it might be possible to investigate this conjecture from
Montonen and Olive duality
\cite{Montonen-Olive,Vafa-Witten} if such a duality
of noncommutative version exists.
(See also \cite{Fucito-Morales-Poghossian}.)
For example, if we assume that the partition function takes the form as 
\begin{eqnarray}
\tilde{Z}_{{\cal N}=4, \tau}^{4dim} =
 \sum_{d=1}^{\infty} \frac{1}{d^2} 
e^{2\pi i \tau k(d)} \ ,
\end{eqnarray}
where $k(d)$ is a instanton number depending on $d$,
restriction to the modular like form
\begin{eqnarray}
\tilde{Z}_{{\cal N}=4, 1/\tau}^{4dim} = \pm
\left( \frac{\tau}{i} \right)^n \tilde{Z}_{{\cal N}=4, \tau}^{4dim}
\end{eqnarray}
might determine $k(d)$, where $n$ is a some number.
Unfortunately, we do not know how to chose a suitable modular like form 
, and above naive conjecture
$
\tilde{Z}_{{\cal N}=4, \tau}^{4dim} = 
 \sum_{d=0}^{\infty} \frac{1}{d^2} 
e^{2\pi i \tau d} \ 
$ does not satisfy this condition.
Anyway, further investigations are necessary to determine the
contribution of the topological term.

%At last,
%$Z_{{\cal N}=4}^{4dim}$ 
%is generating function of Euler number of ${\cal M}_k$.

%%%%%%%%%%%%%%%%%%%%%%%%%%%%%
%%%%%%%%%%%%%%%%%%%%%%%%%%%%%
%%%
%%% 05-11-20
%%%
%%%%%%%%%%%%%%%%%%%%%%%%%%%%%
%%%%%%%%%%%%%%%%%%%%%%%%%%%%%
%Before the end of this section, it is not a waste to
%consider the theory with topological terms.
%In the above discussions, 
At the end of this section, 
we consider the dimensional reduction of the theory with topological
terms. 
In the discussions in section \ref{upf}, 
we use cohomological
field theory without topological terms
like $\int F \wedge F$, and some of them are not supersymmetric
gauge theories in the meaning of the usual supersymmetry.
Now, let us consider the case including topological terms.
%(See also section \ref{modulispace} .)
As an example, let us consider the 4 dimensional case
whose action is given by cohomological terms
and instanton number ;
$ S= \int tr \ \hat{\delta} \Psi + (i \vartheta / 8\pi^2) \int tr \ F \wedge F$.
Let us consider perturbation around classical back ground 
fixed by instanton number, i.e.
$A_{\mu}= A_{\mu}^{(k)} + \delta A_{\mu}^{(k)} \ ,\  
\int tr \ F(A^{(k)}) \wedge F(A^{(k)} )= 8 \pi^2 k $. 
The partition function is given by
$$
Z^{4dim}= \sum_k e^{2\pi i \tau k } \int {{\cal D} \delta A^{(k)}} \dots 
e^{-\int tr\ \hat{\delta} \Psi_k } \ ,
$$
where ${{\cal D} \delta A^{(k)}} \dots $ is the path integral 
measure of the all fields and the functional $\Psi_k$ depends on 
both $A_{\mu}^{(k)} $ and $ \delta A_{\mu}^{(k)}$.
The BRS transformations are induced from (\ref{N=2BRS}) and so on as
$ \hat{\delta} ( \delta A_{\mu}^{(k)} ) = \psi_{\mu}^{(k)} $ etc.
So we have
$$
Z^{4dim} = \sum_k e^{2\pi i \tau k } Z^{4dim }_k \ ,
$$
where $Z^{4dim }_k$ is the perturbative partition function of the $4$
dimension theory without the topological term.
%%%%%%%%%
The action ${\hat \delta} \Psi_k$ is still given by a BRS exact term.
The arguments for the $\theta$-shift invariance of the path integral are valid for $\int {{\cal D} \delta A^{(k)}} \dots 
e^{-\int tr\ \hat{\delta} \Psi_k }$.
Then the dimensional reduction
of the perturbative partition functions
arises at the large $\theta$ limit ;
\begin{equation}
Z^{4dim }_k = Z^{2dim }_k = Z^{0dim }_k \ ,
\label{notsumk}
\end{equation}
where $Z^{2dim }_k$ and $Z^{0dim }_k$ are possible to be described
by partition functions of $2$ and $0$ dimension field theories,
respectively.
Therefore, we find that the universality of the perturbative partition
functions $Z^{4dim }_k$, $Z^{2dim }_k$ and $Z^{0dim }_k$, similar to the claim in section \ref{upf}.

%It is expected that the instanton number is independent from the noncommutative
%parameter $\theta$
%(see for example \cite{sako3,sako4}).
% Therefore, this partition function is still independent
%from the noncommutative parameter,
%if the partition function defined by the action without the topological number is independent from the noncommutative parameter.
%Then the dimensional reduction
%for the path integral $\int {{\cal D} \delta A^{(k)}} \dots 
%e^{-\int tr\ \hat{\delta} \Psi_k }$
%arises at the large $\theta$ limit ;
%\begin{equation}
%Z^{4dim } =
%\sum_k e^{2\pi i \tau k } Z^{2dim }_k =
%\sum_k e^{2\pi i \tau k } Z^{0dim }_k ,
%\label{sumk}
%\end{equation}
%where $Z^{2dim }_k$ and $Z^{0dim }_k$ are possible to be described
%by partition functions of 2 and 0 dimension field theories,
%respectively.
%Seen in this light, above claim is valid 
%for the cases with topological terms.

Now let us discuss the possibility that 
(\ref{notsumk}) means a universality of the partition
functions of usual supersymmetric theories in various dimensions.
Consider the weighted sum of $Z^{4dim }_k$, $Z^{2dim }_k$
and $Z^{0dim }_k$ with weight $e^{2\pi i \tau k }$,
\begin{equation}
\sum_k e^{2\pi i \tau k } Z^{4dim }_k =
\sum_k e^{2\pi i \tau k } Z^{2dim }_k =
\sum_k e^{2\pi i \tau k } Z^{0dim }_k \ .
\label{sumk}
\end{equation}
$\sum_k e^{2\pi i \tau k } Z^{4dim }_k$ is equal to $Z^{4dim }$, the partition of
the $4$ dimension supersymmetric theory including the topological term.
On the other hand, the meanings of the weighted sums $\sum_k e^{2\pi i \tau k } Z^{2dim
}_k$ and $\sum_k e^{2\pi i \tau k } Z^{0dim }_k$ are obscure.
It is unclear that they have the meaning of the partition functions of some lower dimension
theories.
If they can be interpreted as the partition functions of some supersymmetric theories in
lower dimensions, (\ref{sumk}) means a universality of the partition
functions of supersymmetric theories in various dimensions.
To answer whether this statement is true or not,
we need to clarify the following questions. \\
(i) Is the number $k$ expressed in terms of lower dimension theories? \\
(ii) Is the number $k$ interpreted as a topological invariant? And does
it characterize classical solutions of the lower
dimension theories? \\
(iii) Is the total action, the sum of the action defining $Z_k^{2dim }$
or $Z_k^{0dim }$ and the action giving the number $k$,
 equivalent to a supersymmetric action in lower
dimension? \\
At this time, we can only make a few comments on question (i).
We calculated the large $\theta_2$ limit of the
elongated N.C. $U(1)$ $k$-instanton,
that is the reduction from $4$ dimension to $2$ dimension of the
solution.
(For construction of the elongated N.C. $U(1)$ k-instanton, see \cite{IKS-U1}.)
For this case, we can show that $k$ is expressed in terms of $2$
dimension theories, 
\begin{equation}
\theta_1 Tr F_{z_1,{\bar z}_1} = -k \ .
\label{k2dim}
\end{equation}
It may be that 
this fact implies that the number $k$ is expressed in terms of lower
dimension theories.
However, we have no concrete answer to question (ii) and question (iii)
at this time.

%Note that geometrical informations like instanton gauge connections or topological numbers given by integrals of the characteristic classes and so on are reinterpreted in each individual case.
%Space integrations $\int {\cal O}$ are replaced by $Tr_{\cal H} {\cal O}$ in the operator formalism. (See appendix \ref{fock}.)
%If we use the Fock space as this Hilbert space, this trace operation is
%the sum of the diagonal elements $\langle n_1 , n_2| {\cal O} |n_1 , n_2
%\rangle$ in the 4 dimensional theory.
%After dimensional reduction, $Tr_{\cal H} {\cal O}$ is regarded as
%the sum of the diagonal elements $\langle n| {\cal O} |n \rangle $
%in the 2 dimensional theory or 
%formal trace of infinite dimensional matrix model in the 
%0 dimensional theory.
%{}From this correspondence, space integrals, topological invariants
%or some geometrical informations in the four dimension are reinterpreted 
%as ones in lower dimensions.

%%%%%%%%%%%%%%%%%%%%%%%%%%%%%%%
%%%%%%%%%%%%%%%%%%%%%%%%%%%%%%%
%%%%%%%%%%%%%%%%%%%%%%%%%%%%%%%
\section{Conclusions and Discussions} \label{conclusion}

We investigated  
cohomological gauge theories in N.C. ${\mathbb R}^{2D}$.
We saw that vacuum expectation values of the theories 
do not depend on noncommutative parameters, and
the large noncommutative parameter limit 
is equivalent to the dimensional reduction.
As a result of these facts, we showed that 
two types of cohomological theories defined in N.C.
${\mathbb R}^{2D}$ and N.C. ${\mathbb R}^{2D+2}$
are equivalent, if they are connected through dimensional
reduction.
Therefore, we found several partition functions of noncommutative supersymmetric
Yang-Mills theories in various dimensions are equivalent, 
when they are connected
by dimensional reduction from $2+2D$ to $2D$.
Using this technique and requiring some natural assumptions, 
we determine the 
partition function of the ${\cal N}=4$ U(1) gauge theory 
in N.C. ${\mathbb R}^4$, where the action
does not include the topological term $\tau \int F \wedge F$,
and the result is equivalent to the partition function
of ( 8dim , ${{\cal N}=2}$ ) , ( 6dim , ${{\cal N}=2}$ ) ,
( 2dim , ${{\cal N}=8}$ ) and the IKKT matrix model
given by their dimensional reduction
to 0 dim.
%These results are consistent with the 
%D-brane picture of the Type II string.
The case including the topological term was discussed, too.
\\

Let us list some left problems below.
In this article, concrete partition functions
are given for the ${\cal N}=4$ U(1) gauge theory 
in N.C. ${\mathbb R}^4$ and the series
connecting to it by dimensional reduction.
So, we are interested in
 N.C.non-abelian cases.
To calculate them, we have to find some new formulation like
MNS, because we know
the partition function concerning $su(N)$ but
we need it for $su(N)\times su(M)$ for $U(M)$ theory.

Next, we had qualitative observation of
${\cal N}=2$ 4-dim case but we do not do
quantitative approach.
So, we have to do the more detail analysis for the ${\cal N}=2$
super Yang-Mills cases.
We saw in section \ref{modulispace},
after taking large $\theta$ limit, moduli space
is described by Monads in ${\cal N}=4$ 4-dim case.
{}From the analogy with ${\cal N}=4$ 4-dim case,
direct and smooth connections between noncommutative instanton moduli spaces
and ADHM spaces might be given in ${\cal N}=2$ 4-dim case.
%by adding bi-fundamental
%matter field to the ${\cal N}=2$
%super Yang-Mills theory.

Other important problems are applications to the various fuzzy spaces,
$T^d_{\theta}$ , ${\mathbb CP}^d_N$, and so on.
Since these noncommutative spaces are 
expressed by finite dimensional Hilbert spaces,
the dimensional reduction will not occur
at the large $\theta$ limit
despite omitting kinetic terms.

Wide spread applications of the technology
of this article are going to happen in many cases other 
than above subjects.
All of them are left for future works.\\

\noindent
{\bf Acknowledgements}\\
We would like to thank Takuya Miyazaki for useful remarks and discussion.
Discussions during the workshop ``An International Meeting
Noncommutative Geometry, 
K-theory and Physics 2005"
were useful to complete this work.

%%%%%%%%%%%%%%%%%%%%%%%%%%%%%%%%%%%%%%%%%%%%%%%%%%%%%%%%%%%%%%%
%%%%%%%%%%%%%%% Appendix %%%%%%%%%%%%%%%%%%%%%%%%%%%%%%%%%%%%%%
%%%%%%%%%%%%%%%%%%%%%%%%%%%%%%%%%%%%%%%%%%%%%%%%%%%%%%%%%%%%%%%

%%%
%%%%%%%%
\appendix
\section{Fock Space}\label{fock}
Let us consider N.C. ${\mathbb R}^{2D}$.
First of all, we introduce following operators,
\begin{eqnarray}
a_{i} \equiv \frac{z_{i}}{\sqrt{\theta^{2i-1,2i}}} &,& z_{i}\equiv\frac{1}{\sqrt{2}}(x^{2i-1}+ix^{2i}) , \nonumber \\
a_{i}^{\dag} \equiv \frac{\bar{z}_{i}}{\sqrt{\theta^{2i-1,2i}}} &,& \bar{z}_{i}\equiv\frac{1}{\sqrt{2}}(x^{2i-1}-ix^{2i}) ,
\label{acopa}
\end{eqnarray}
where $i$ runs from $1$ to $D$, and $a_{i}$ and
$a_{i}^{\dag}$ satisfy 
\begin{equation}
[a_{i},a_{j}^{\dag}] = \delta_{i j} .
\label{cra}
\end{equation}
We often use the symbol $\theta^i$ defined as
\begin{equation}
\theta^i = + \theta^{2i-1,2i} = - \theta^{2i,2i-1} .
\label{thetaia}
\end{equation}

The Hilbert space is constructed as the Fock space,
\begin{eqnarray}
 {\cal H}&=&\bigoplus {\bb C}\left|n_1,\cdots,n_D\right>\;,
\nonumber\\
 \left|n_1,\cdots,n_D\right>&\equiv&
  \frac{(a_1^{\dag})^{n_1}\cdots(a_D^{\dag})^{n_D}}
  {\sqrt{n_1!\cdots n_D!}}\left|0,\cdots,0\right>\;.
\label{fockspa}
\end{eqnarray}
$a_{i}$ and $a_{i}^{\dag}$ operate on
$\left|n_1,\cdots,n_D\right>$ as follows
\begin{eqnarray}
 a_{i}\left|n_1,\cdots,n_D\right>
  &=&\sqrt{n_{i}}\left|n_1,\cdots,n_{i}-1,\cdots,n_D\right>\;,
\nonumber \\
 a_{i}^{\dag}\left|n_1,\cdots,n_D\right>
  &=&\sqrt{n_{i}+1}\left|n_1,\cdots,n_{i}+1,\cdots,n_D\right>\;.
\label{aonna}
\end{eqnarray}
$\left|n_1,\cdots,n_D\right>$ are the eigenstates of the number operator
$\hat{n}_{i}\equiv a_{i}^{\dag}a_{i}$,
\begin{equation}
\hat{n}_{i} \left|n_1,\cdots,n_D\right> = n_i \left|n_1,\cdots,n_D\right> .
\label{nopa}
\end{equation}
Arbitrary operator has following expression;
\begin{equation}
 \hat{\cal O}=\sum_{n_1,m_1}\cdots\sum_{n_D,m_D}
  {\cal O}_{m_1\cdots m_D}^{n_1\cdots n_D}
  \left|n_1,\cdots,n_D\right>\left<m_1,\cdots,m_D\right|\;.
\label{fockexa}
\nonumber
\end{equation}

Let us consider $2D=4$ case. The Hilbert space ${\cal H}$ is
expanded by the Fock basis $|n_1 , n_2 \rangle$,
\begin{eqnarray}
{\cal H} &=& \bigoplus {\mathbb C} |n_1 , n_2 \rangle , \nonumber \\
|n_1 , n_2 \rangle &=& \frac{({ a}^\dagger_1)^{n_1} ({ a}^\dagger_2)^{n_2}}{\sqrt{n_1 ! n_2 !}} |0 , 0 \rangle .
\label{focksp2a}
\end{eqnarray}
$a^\dagger_i$ and $a_i$ are expressed as
\begin{eqnarray}
{ a}^\dagger_1 = \sum_{n_1 =0}^{\infty} \sqrt{n_1 +1} |n_1 +1 , n_2 \rangle \langle n_1 , n_2| &,& { a}^\dagger_2 = \sum_{n_2 =0}^{\infty} \sqrt{n_2 +1} |n_1 , n_2 +1 \rangle \langle n_1 , n_2| , \nonumber \\
{ a}_1 = \sum_{n_1 =0}^{\infty} \sqrt{n_1 +1} |n_1 , n_2 \rangle \langle n_1 +1 , n_2| &,& { a}_2 = \sum_{n_2 =0}^{\infty} \sqrt{n_2 +1} |n_1 , n_2 \rangle \langle n_1 , n_2 +1| . \nonumber \\
\label{aa} 
\end{eqnarray}

The finite dimensional truncation ${\cal H}_N$ can be defined by several
ways. 
One definition of ${\cal H}_N$ is given by
\begin{equation}
{\cal H}_{N} = \bigoplus_{n_1 =0 , n_2 =0}^{n_1 =N_c , n_2 =N_c} {\mathbb C} |n_1 , n_2 \rangle ,
\label{hn1a}
\end{equation}
where $N_c$ is a finite integer number.
By the definition, we obtain
\begin{equation}
\mbox{dim. of} \ \ {\cal H}_N = (N_c +1)^2 = N ,
\label{d1a}
\end{equation}
and 
\begin{equation}
{\bf 1}_N = \sum_{n_1 =0 , n_2 =0}^{n_1 =N_c , n_2 =N_c} |n_1 , n_2 \rangle \langle n_1 , n_2| .
\label{11a}
\end{equation}

Another definition of ${\cal H}_N$ is given by
\begin{equation}
{\cal H}_{N} = \bigoplus_{n_1 =0 , n_2 =0}^{n_1 + n_2 =N_c} {\mathbb C} |n_1 , n_2 \rangle .
\label{hn2a}
\end{equation}
In this case, 
\begin{equation}
\mbox{dim. of} \ \ {\cal H}_N = \frac{(N_c +1)(N_c +2)}{2} = N ,
\label{d2a}
\end{equation}
and
\begin{equation}
{\bf 1}_N = \sum_{n_1 =0 , n_2 =0}^{n_1 + n_2 =N_c} |n_1 , n_2 \rangle \langle n_1 , n_2| .
\label{12a}
\end{equation}

By using the definition of ${\bf 1}_N$, (\ref{11a}) or (\ref{12a}), and
the following expressions of the differential operators ${\hat \partial}_\mu$
in terms of $a^\dagger_i$ and $a_i$,
\begin{eqnarray}
{\hat \partial}_1 = \frac{1}{\sqrt{2\theta^1}} (a_1 - a^\dagger_1) & , & {\hat \partial}_2 = \frac{-i}{\sqrt{2\theta^1}} (a_1 + a^\dagger_1) , \nonumber \\
{\hat \partial}_3 = \frac{1}{\sqrt{2\theta^2}} (a_2 - a^\dagger_2) & , & {\hat \partial}_4 = \frac{-i}{\sqrt{2\theta^2}} (a_2 + a^\dagger_2) ,
\label{daa}
\end{eqnarray}

Given the definition of ${\cal H}_N$, for example by (\ref{hn1}),
we obtain
\begin{eqnarray}
 & & [a_1 , {\bf 1}_N] = -\sqrt{N+1} \sum_{n_2=0}^{N} |N,n_2 \rangle \langle N+1,n_2| , \nonumber \\
 & & [a_1^{\dagger} , {\bf 1}_N] = +\sqrt{N+1} \sum_{n_2=0}^{N} |N+1,n_2 \rangle \langle N,n_2| , \nonumber \\
 & & [a_2 , {\bf 1}_N] = -\sqrt{N+1} \sum_{n_1=0}^{N} |n_1,N \rangle \langle n_1,N+1| , \nonumber \\
 & & [a_2^{\dagger} , {\bf 1}_N] = +\sqrt{N+1} \sum_{n_1=0}^{N} |n_1,N+1 \rangle \langle n_1,N| .
\label{a1a}
\end{eqnarray}
{}From (\ref{a1a}) and (\ref{daa}), we obtain
\begin{eqnarray}
T_1 = \frac{1}{\sqrt{2\theta^1}} & ( & - \sqrt{N+1} \sum_{n_2=0}^{N} |N,n_2 \rangle \langle N+1,n_2| \nonumber \\
& & - \sqrt{N+1} \sum_{n_2=0}^{N} |N+1,n_2 \rangle \langle N,n_2| \ \ \ \ \ ) , \nonumber \\
T_2 = \frac{-i}{\sqrt{2\theta^1}} & ( & - \sqrt{N+1} \sum_{n_2=0}^{N} |N,n_2 \rangle \langle N+1,n_2| \nonumber \\
& & + \sqrt{N+1} \sum_{n_2=0}^{N} |N+1,n_2 \rangle \langle N,n_2| \ \ \ \ \ ) , \nonumber \\
T_3 = \frac{1}{\sqrt{2\theta^2}} & ( & - \sqrt{N+1} \sum_{n_1=0}^{N} |n_1,N \rangle \langle n_1,N+1| \nonumber \\
& & - \sqrt{N+1} \sum_{n_1=0}^{N} |n_1,N+1 \rangle \langle n_1,N| \ \ \ \ \ ) , \nonumber \\
T_4 = \frac{-i}{\sqrt{2\theta^2}} & ( & - \sqrt{N+1} \sum_{n_1=0}^{N} |n_1,N \rangle \langle n_1,N+1| \nonumber \\
& & + \sqrt{N+1} \sum_{n_1=0}^{N} |n_1,N+1 \rangle \langle n_1,N| \ \ \ \ \ ) .
\label{T1234a}
\end{eqnarray}
Using (\ref{11a}) and (\ref{T1234a}), we can show
%\begin{equation}
%Tr_{\cal H} \ T_\mu T_\nu = \frac{N}{\theta^i} \delta_{\mu \nu} .
%\label{trexo}
%\end{equation}
%${\cal T}_\mu$ is defined by
%\begin{equation}
%{\cal T}_\mu = \frac{\sqrt{\theta^i}}{\sqrt{N}} T_\mu ,
%\label{calT}
%\end{equation}
\begin{equation}
Tr_{\cal H} \ {\bf 1}_N \ {\bf 1}_N = N \ \ , \ \ Tr_{\cal H} \ T_\mu \ T_\nu = + \frac{1}{\theta^i} N \delta_{\mu \nu} \ \ , \ \ Tr_{\cal H} {\bf 1}_N \ T_\mu = 0 .
\label{tr2oa}
\end{equation}
%where
%\begin{equation}
%T_\mu = [{\hat \partial}_\mu , {\bf 1}_N] .
%\label{Ta}
%\end{equation}
Also, we can obtain
\begin{eqnarray}
Tr_{\cal H} {\bf 1}_N [{\hat \partial}_\mu , {\bf 1}_N]= 0 &,&
Tr_{\cal H} {\bf 1}_N [{\hat \partial}_\mu , T_\nu] = - \frac{N}{\theta^i} \delta_{\mu \nu} , \nonumber \\
Tr_{\cal H} T_\mu [{\hat \partial}_\nu , {\bf 1}_N] = + \frac{N}{\theta^i} \delta_{\mu \nu} &,& Tr_{\cal H} T_\mu [{\hat \partial}_\nu , T_\rho ] = 0 \ \ \ ,
\label{tr2doa}
\end{eqnarray}
and
\begin{eqnarray}
Tr_{\cal H} {\bf 1}_N [{\bf 1}_N , {\bf 1}_N] = 0 &,& 
Tr_{\cal H} {\bf 1}_N [{\bf 1}_N , T_\mu] = 0 , \nonumber \\ 
Tr_{\cal H} {\bf 1}_N [T_\mu , T_\nu] = + i N \theta^{-1}_{\mu \nu} &,& 
Tr_{\cal H} T_\mu [T_\nu , T_\rho] =0 .
\label{tr3oa}
\end{eqnarray}

Let us define ${\cal I}$ and ${\cal T}_\mu$ as,
\begin{equation}
{\cal I} = \frac{1}{\sqrt{N}} {\bf 1}_N ,
\label{calIa}
\end{equation}
and
\begin{equation}
{\cal T}_\mu = \frac{\sqrt{\theta_i}}{\sqrt{N}} T_\mu .
\label{calTa}
\end{equation}
By definition, 
\begin{equation}
{\cal T}_\mu = \sqrt{\theta_i} [{\hat \partial}_\mu , {\cal I}] . 
\label{calTIa}
\end{equation}

Using ${\cal I}$ and ${\cal T}_\mu$,
(\ref{tr2oa}),(\ref{tr2doa}) and (\ref{tr3oa}) are rewritten into
\begin{equation}
Tr_{\cal H} \ {\cal I} \ {\cal I} = 1 \ \ , \ \ Tr_{\cal H} \ {\cal T}_\mu \ {\cal T}_\nu = \delta_{\mu \nu} \ \ , \ \ Tr_{\cal H} \ {\cal I} \ {\cal T}_\mu = 0 ,
\label{tr2a}
\end{equation}
\begin{eqnarray}
Tr_{\cal H} {\cal I} [{\hat \partial}_\mu , {\cal I}]= 0 &,&
Tr_{\cal H} {\cal I} [{\hat \partial}_\mu , {\cal T}_\nu] = - \frac{1}{\sqrt{\theta^i}} \delta_{\mu \nu} , \nonumber \\
Tr_{\cal H} {\cal T}_\mu [{\hat \partial}_\nu , {\cal I}] = + \frac{1}{\sqrt{\theta^i}} \delta_{\mu \nu} &,& Tr_{\cal H} {\cal T}_\mu [{\hat \partial}_\nu , {\cal T}_\rho ] = 0 \ \ \ ,
\label{tr2da}
\end{eqnarray}
and
\begin{eqnarray}
Tr_{\cal H} {\cal I} [{\cal I} , {\cal I}] = 0 &,& 
Tr_{\cal H} {\cal I} [{\cal I} , {\cal T}_\mu] = 0 , \nonumber \\ 
Tr_{\cal H} {\cal I} [{\cal T}_\mu , {\cal T}_\nu] = + \frac{i
 \theta^i}{\sqrt{N}} \theta^{-1}_{\mu \nu} &,& 
Tr_{\cal H} {\cal T}_\mu [{\cal T}_\nu , {\cal T}_\rho] =0 .
\label{tr3a}
\end{eqnarray}

The same formulae as (\ref{tr2a}), (\ref{tr2da}) and (\ref{tr3a}) hold
for the case of (\ref{hn2a}).
The difference between the definitions
of ${\cal H}_{N}$'s, (\ref{hn1a}) and (\ref{hn2a}), are absorbed in
$\mbox{dim. of} \ {\cal H}_N$.

It is worthwhile to notice that the independence of the precise
definitions of ${\cal H}_N$ holds generally.
The proof is done by using the discrete version of Stokes's theorem for
the boundary of the finite truncated Fock space \cite{sako3,sako4}.

%%%%%%%%%%%%%%%%%%%%%%%%%%%%%%%%%%%%%%%%%%%%%%%%%%%%
%%%%%%%%%%%%%%%%%%%%%%% REV %%%%%%%%%%%%%%%%%%%%%%

\section{ Large $\theta$ limit} \label{B}

In this article, we removed the terms including 
${\partial}_{\mu} = -i \theta^{-1}_{\mu \nu}[x^{\nu}\ ,\ * \ ]$
in the lagrangian when we calculated the partition function
without zero mode integrals in the large $\theta$ limit.
If we consider some specific fixed function $f(x)$,
then expression of
${\partial}_{\mu} f(x)= -i \theta^{-1}_{\mu \nu}[x^{\nu}\ , f(x)  ]$
is not changed by taking large $\theta$ limit 
because $[x^{\nu}\ , f(x) ]$ becomes large 
with $\theta$.
Therefore, someone might think that the process of removing terms including 
${\partial}_{\mu}$ is not correct.
However, we have to recall that our lagrangian is changed
by $\theta$ variation and then the equations of motion and
BPS equations are changed.
Then the solutions 
of the equations , which make much contribution to 
the partition functions and vacuum expectation values,
are changed by $\theta$ changing.
It follows that the terms including derivatives become irrelevant. 
In this section, we show concretely the validity
of taking the terms including 
${\partial}_{\mu} = -i \theta^{-1}_{\mu \nu}[x^{\nu}\ ,\ * \ ]$
away from lagrangians at the large $\theta$ limit.\\

The BPS eqs. in this paper are given by
differential equations of first order ;
\begin{eqnarray}
 \sum_{i,I} c_{i I , k} \partial_{z_i} f_I + V_k(f_J) = 
 0 \ , \label{diffeq}
\end{eqnarray}
where $ f_I $ are fields, $V_k(f_I)$ are some quadratic polynomial 
in $ f_I $ and $c_{i I ,k} $ are some constants.
$k = 1, \dots , n$ , where $n$ is the number of elements of $f_I$
minus degree of gauge freedom.
For example, BPS eqs. of ${\cal N}=4$ 4-dim. gauge theory are
\begin{eqnarray}
F^{+\mu\nu}-i[B^+_{\mu\rho},{B^+_{\nu}}^{\rho}] 
-i[B^+_{\mu\nu},c]=0 \; , \;
2D^\mu B^+_{\mu\rho}
+D_\rho c =0 \ .
\end{eqnarray}
Let us consider (\ref{diffeq}) by using the Fock basis 
;
\begin{eqnarray}
B_k (\hat{f}_I , \theta )\equiv c_{i I , k}^{+} \frac{1}{\sqrt{\theta^i}}
[ a_i , \hat{f}_I ] 
+ c_{i I , k}^{-} \frac{1}{\sqrt{\theta^i}}
[ a_i^{\dagger} , \hat{f}_I ] 
+ V_k(\hat{f}_J) =d_{i, k} \frac{1}{\theta^i} \ . \label{diffeq_fock}
\end{eqnarray}
Here $d_{i ,k} \frac{1}{\theta^i} $ are constants derived from 
$ [\partial_{z_i} , \partial_{\bar{z}_i}] $.
For example, eqs. of ${\cal N}=4$ 4-dim. cases are given by
\begin{eqnarray}
P^+_{\mu \nu \rho \tau} [\hat{D}^{\rho} , \hat{D}^{\tau} ] 
+[B^+_{\mu\rho},{B^+_{\nu}}^{\rho}] 
+[B^+_{\mu\nu},c ]&=&i
\Big( P^+_{\mu \nu \rho \tau} (\theta^{-1})^{\rho \tau} \Big)\; , 
\label{4dBPS_1} \\
2[\hat{D}^{\mu} , B_{+\mu\rho} ]
+[\hat{D}_{\rho} , c ] &=& 0 \ , \label{4dBPS_2}
\end{eqnarray}
where $P^+_{\mu\nu \rho \tau}$ is self-dual projection operator
and $\hat{D}_{\mu} = \hat{\partial}_{\mu} + i {A}_{\mu}$.
When we take $\theta^{\mu \nu}$ as (\ref{theta1}),
the righthand side of (\ref{4dBPS_1}) is rewritten as
\begin{eqnarray}
P^+_{\mu \nu \rho \tau} (\theta^{-1})^{\rho \tau}
&=& - \frac{\varepsilon_{\mu \nu}}{2} \left(
\frac{1}{\theta^1}+ \frac{1}{\theta^2} \right) , \nonumber \\
\Big( \varepsilon_{\mu \nu} \Big ) & \equiv &
\left(
\begin{array}{cc|cc}
0&1& & \\
-1&0 & & \\
\hline
 & & 0 &1 \\
 & & -1&0
\end{array}
\right) . \nonumber 
\end{eqnarray}
$\hat{f_I}$ is a operator representation of $f_I$,
i.e. $\hat{f_I}= \sum (f_I)_{n_1, \dots , n_D}^{m_1, \dots m_D}
|n_1, \dots , n_D \rangle \langle m_1, \dots m_D |$.
In this representation, the BPS eqs. are just simultaneous
quadratic equations, and the noncommutative parameters
$\theta^i$ appear in only first 2 terms and right hand side
in (\ref{diffeq_fock}).
Note that solutions of (\ref{diffeq_fock}) depend on $\theta^i$
but variables $(f_I)_{n_1, \dots , n_D}^{m_1, \dots m_D}$ themselves 
do not depend on $\theta$.
For this reason, BPS equations are truncated to
\begin{eqnarray}
B_k (\hat{f}_I , \infty )\equiv
V_k(\hat{f}_J) =0 \ , \label{BPSeq_fock_largetheta}
\end{eqnarray}
at the $\theta \rightarrow \infty$ limit.
Such truncations have been discussed in many works,
see for example \cite{GMS,G-H-S,Jatkar-Mandal-Wadia}.
Thus, it becomes clear that
terms including 
${\partial}_{\mu} = -i \theta^{-1}_{\mu \nu}[x^{\nu}\ ,\ * \ ]$
in the lagrangian 
become irrelevant
at the large $\theta$ limit. \\

However, the above discussion is insufficient for the proof
which justifies removing terms including 
${\partial}_{\mu}$.
Because we assume the convergency of path integral 
which has not been confirmed
when we formally prove that partition functions and 
vacuum expectation values of observables 
do not depend on $\theta$.
%The formal proof is valid if path integral is enough
%convergent, the detail of convergency will be seen soon,
%then 
Therefore, we have to check our models satisfying the convergency conditions.
To understand this statement, let us consider the following example.

Let $f_i$ be dynamical variables and assume that action functional 
take the following form :
\begin{eqnarray}
S_{\epsilon} [f] = S_0 + \epsilon S_1 \ ,
\end{eqnarray}
where $\epsilon$ is a some constant,
$S_0$ and $S_{\epsilon}$ are BRS exact actions,
and they do not depend on $\epsilon$.
Let us expand the partition function as
\begin{eqnarray}
Z_{\epsilon} &=& \int {\cal D}f e^{-S_{\epsilon}}   \label{p_f_epsilon} \\
&=& \int {\cal D}f \ e^{-S_0} (1- \epsilon S_1 +\frac{1}{2} \epsilon^2 S_1^2- \cdots )
\end{eqnarray}
and introduce
\begin{eqnarray}
Z_0 \equiv \int {\cal D}f e^{-S_0} \label{p_f_0} \ .
\end{eqnarray}
If $e^{-S_0}$ damp the integrand, the integral 
\begin{eqnarray}
\int {\cal D}f e^{-S_0} \epsilon^n S_1^n \ , \ \mbox{for} \ n\ge 1
\label{terms}
\end{eqnarray}
is well defined. Then, $Z_{\epsilon}$ does not
depend on $\epsilon$, i.e.
$$
Z_{\epsilon} = Z_0  \ ,
$$
because 
$S_1^n $ is a BRS exact term and 
$$
\int {\cal D}f e^{-S_0} \epsilon^n S_1^n = 0 \ , \ \mbox{for} \ n\ge 1 \ .
$$
%Therefore, convergency of (\ref{p_f_0}) and 
%the fact that $e^{-S_0}$ damp integrands 
%are necessary
%for the partition function being independent of $\epsilon$.
Therefore, we found that we have to verify that
(\ref{p_f_0}) is well defined and
$e^{-S_0}$ damp integrands for proof of $\epsilon$ independence.

To get a feeling for how all of this should work out,
consider simple toy models.
At first, let us consider the toy model given by Vafa and Witten
in section 2 of \cite{Vafa-Witten}.
Let
$x , y , H_1 $ and $H_2$ be real bosonic variables, and 
$\psi_x , \psi_y , \chi_1$ and $\chi_2$ be fermionic variables.
We define BRS transformations by
\begin{eqnarray}
\hat{\delta} x = \psi_x \ , \ 
\hat{\delta} y= \psi_y \ , \ 
\hat{\delta} \chi_1 = H_1 \ , \ 
\hat{\delta}  \chi_2 = H_2\ .
\end{eqnarray}
Consider following action
\begin{eqnarray}
S^{toy1}_{\epsilon} &=&  \hat{\delta}
\{ \chi_1 ( H_1 + 2i ( x^2 - \epsilon - y^2) )
+\chi_2 ( H_2 + 2i ( 2xy ))
\} \ , \nonumber \\
&=& S_0^{toy1} + \epsilon S_1^{toy1}  \ ,
\end{eqnarray}
where
\begin{eqnarray}
S_0^{toy1} &=&  \hat{\delta}
\{ \chi_1 ( H_1 + 2i ( x^2  - y^2) )
+\chi_2 ( H_2 + 2i ( 2xy )) 
\} \ , \nonumber \\
S_1^{toy1} &=&  \hat{\delta}  \chi_1 \ . \nonumber
\end{eqnarray}
$e^{-S_0^{toy1}}$ makes the integral 
\begin{eqnarray}
\int {\cal D}f e^{-S_0^{toy1}} \epsilon^n (S_1^{toy1})^n \ , 
\ \mbox{for} \ n\ge 1
\label{terms}
\end{eqnarray}
be well defined, and
\begin{eqnarray}
Z_{\epsilon}^{toy1} = Z_0^{toy1} \equiv \int {\cal D}f e^{-S_0^{toy1}}\ .
\label{epsilon0}
\end{eqnarray}
Indeed, we can easily
perform the direct calculations of the partition functions
$Z_{\epsilon}^{toy1}$ and $Z_0^{toy1}$, respectively,
and their results reproduce (\ref{epsilon0}).
Note that degeneracy of the solutions does not affect
the independence of $\epsilon$.
In this case, when $\epsilon \neq 0$ equations are 
given by $ x^2 - \epsilon -y^2 =0$ and $2xy=0$ , then
the solutions are given as $(x,y)= ( \pm \sqrt{\epsilon} , 0 )$.
These two sets of solutions become degenerate in $\epsilon \rightarrow 0$.
Despite such singularities, path integrals moderate them,
and the partition function is smooth at $\epsilon = 0$.

As the second example,
consider the following action
\begin{eqnarray}
S^{toy2}_{\epsilon} &=&  \hat{\delta}
\{ \chi_1 ( H_1 + 2i ( x^2 - \epsilon ) )
+\chi_2 ( H_2 + 2i ( 2xy ))
\} \ , \nonumber \\
&=& S_0^{toy2} + \epsilon S_1^{toy2}  \ ,
\end{eqnarray}
where
\begin{eqnarray}
S_0^{toy2} &=&  \hat{\delta}
\{ \chi_1 ( H_1 + 2i ( x^2  ) )
+\chi_2 ( H_2 + 2i ( 2xy )) 
\} \ , \nonumber \\
S_1^{toy2} &=&  \hat{\delta}  \chi_1 \ . \nonumber
\end{eqnarray}
At first glance, the partition function $Z_{\epsilon}^{toy2}$
looks independent of $\epsilon$ from the formal discussion.
But $e^{-S_0^{toy2}}$ does not damp the integrals
in this case, then $Z_{\epsilon}^{toy2}$ depends on $\epsilon$.
Indeed,
\begin{eqnarray}
Z_{\epsilon}^{toy2} &=&
\int \frac{dx}{\sqrt{2 \pi}} \frac{dy}{\sqrt{2 \pi}} \frac{dH_1}{\sqrt{2
\pi}} \frac{dH_2}{\sqrt{2 \pi}} d\psi_x d\psi_y d\chi_1 d\chi_2 \ 
\exp{\left( -S^{toy2}_{\epsilon} \right)} \nonumber \\
&=& 1 + \frac{\epsilon}{2} \pi^{-1/2} +O(\epsilon^2 ) \ .
\end{eqnarray}
These observations show that we have to check the
convergency of $e^{-S_0}$ where the action $S_0$ is 
$\theta$ independent part of the total action,
before removing terms including $\theta^{-1}$ from action.\\

Let us now attempt to investigate the specific case of
${\cal N}=4$ 4-dim.
First we consider the case of $\theta^1 = -\theta^2$.
This is very special case and we can understand the validity of 
removing the terms including  ${\partial}_{\mu}$ not from 
above discussions but from the following discussions.
Using $\theta^1 = -\theta^2$, the BPS eqs 
(\ref{4dBPS_1}) and (\ref{4dBPS_2}) are replaced by
\begin{eqnarray}
P^+_{\mu \nu \rho \tau} [\hat{D}^{\rho} , \hat{D}^{\tau} ] 
+[B^+_{\mu\rho},{B^+_{\nu}}^{\rho}] 
+[B^+_{\mu\nu},c ]&=& 0\; , 
\label{4dBPS_3} \\
2[\hat{D}^{\mu} , B_{+\mu\rho} ]
+[\hat{D}_{\rho} , c ] &=& 0 \ . \label{4dBPS_4}
\end{eqnarray}
On the contrary, the BPS eqs. of the large $\theta$ limit
are given by
\begin{eqnarray}
-P^+_{\mu \nu \rho \tau} [A^{\rho } , A^{\tau } ] 
+[B^{+ }_{\mu\rho},{B^{+ }_{\nu \tau}}] \delta^{\tau \rho} 
+[B^{+ }_{\mu\nu},c ]&=& 0\; , 
\label{4dBPS_5} \\
2[A^{\mu } , B^{+ }_{\mu\rho} ]
+[A_{\rho} , c ] &=& 0 \ . \label{4dBPS_6}
\end{eqnarray}
(\ref{4dBPS_5} , \ref{4dBPS_6}) are 
equivalent to (\ref{4dBPS_3} , \ref{4dBPS_4}) with $1/\theta = 0$.
The correspondence of these and more general cases are already known in 
\cite{Aoki-Ishibashi-Iso-Kawai-Tada} and \cite{Sochichiu}, that is, we can identify (\ref{4dBPS_3} , \ref{4dBPS_4})
and (\ref{4dBPS_5} , \ref{4dBPS_6}) by redifining 
\begin{eqnarray}
iA_{\mu} = \hat{D}_{\mu} \ . \label{correspond} 
\end{eqnarray}
{}This is a trivial one to one correspondence between 
the large $\theta$ limit and finite $\theta^1 = -\theta^2$ case.
Under change of variables (\ref{correspond}),
the path integral measure does not cause nontrivial
Jacobian, then theories characterized 
by (\ref{4dBPS_3},\ref{4dBPS_4}) and (\ref{4dBPS_5},\ref{4dBPS_6})
are equivalent quantum theories. 
{}From this correspondence, it is clear that we can remove
the terms including  ${\partial}_{\mu}$ from its action without changing.\\

Before investigating $\theta^1 \neq -\theta^2$ case, let us consider 
\begin{eqnarray}
S_{\epsilon} &=& S_0 + \epsilon S_1 \nonumber \\
S_0&=& Tr_{\cal H} tr \ \
 \hat{\delta}_+ \left\{ \chi^+_{\mu\nu}
\left( H^{+\mu\nu}-(
P^+_{\mu \nu \rho \tau} [\hat{D}^{\rho} , \hat{D}^{\tau} ]
+[B^+_{\mu\rho},B^+_{\nu\sigma}] \delta^{\rho\sigma}+
%\right. \right. \nonumber \\
%& & \ \ \ \ \ \ \ \ \ \ \ \ \ \ \ \ \ \ \ \ \ 
[B^+_{\mu\nu},c] ) \right)
\right\}
\nonumber
\\
& &+Tr_{\cal H} tr \ \
\hat{\delta}_+ \{
\chi^{\rho }\left(
H_{\rho}-i(
-2[ \hat{D}^\mu , B_{\mu\rho}^+ ]
-[ \hat{D}_\rho  , c ]
)\right)
\}
\nonumber \\
& & +Tr_{\cal H} tr \ \
 \hat{\delta}_+  \{ i[\phi , {\bar \phi}] \eta - i {\bar \eta} [c , {\bar 
\phi}]
+ i [ B^{+ \ \mu \nu} , {\bar \phi}] \psi^+_{\mu \nu} +
 ([ \hat{D}_\mu  , {\bar \phi} ]) \psi_{\mu}  \}  
\label{S_e} \\
%&=& (\hbox{Twisted {\cal N}=4 Super Y-M} )
S_1 &=& i \chi_{\mu \nu}^+ 
\varepsilon^{\mu \nu} , \nonumber 
\end{eqnarray}
and their partition functions :
\begin{eqnarray}
Z_{{\cal N}=4,\epsilon} = \int {\cal D} f e^{-S_{\epsilon}} \ \ , \ \ 
Z_{{\cal N}=4,0} = \int {\cal D} f e^{-S_{0}} \ .
\end{eqnarray}
Note that $S_0$ is equivalent to the action of the 
Yang-Mills theory of $\theta^1 = -\theta^2$
and IKKT matrix model when its gauge group is U(1).
Therefore, it is natural to assume that $\exp (-S_0)$
damp the path integral of an arbitrary observable.
Indeed, this assumption is required in MNS too
\cite{Moore-Nekrasov-Shatashvili}.
{}From above discussion and this assumption,
we can conclude that
\begin{eqnarray}\label{e_independence}
Z_{{\cal N}=4,\epsilon} = Z_{{\cal N}=4,0} .
\end{eqnarray}

Next step, we consider $\theta^1 \neq -\theta^2$ case. 
Its action is equivalent to (\ref{S_e}) if 
$$ \epsilon = -\frac{1}{2} 
\left( \frac{1}{\theta^1} + \frac{1}{\theta^2} \right) \ .$$
Under the above assumption that $\exp (-S_0)$
damp integrands of path integrals, as we saw in
(\ref{e_independence}), $Z_{{\cal N}=4,\epsilon} $ does not
depend on $\epsilon$ .
Therefore, the partition function of $\theta^1 \neq -\theta^2$ case
is equal to the partition function of $\theta^1 = -\theta^2$
whose BPS eqs. are given by (\ref{4dBPS_3} , \ref{4dBPS_4}),
furthermore the partition function is equal to
the partition function whose action functional is given by removing
derivative terms and its BPS eqs. are given by
(\ref{4dBPS_5} , \ref{4dBPS_6}).
\\

In the above discussion, 
we have closely studied the case of dimensional reduction from
${\cal N}=4$ 4-dim. to 0-dim.
But it is clear that we can apply the above general
discussion to other dimensional cases or the cases of the
${\cal N}=2$ 4-dim. model and the series given by its dimensional reduction.  
All these things make it clear that
it is proper procedure to
remove the terms including 
${\partial}_{\mu} = -i \theta^{-1}_{\mu \nu}[x^{\nu}\ ,\ * \ ]$
from lagrangians at the large $\theta$ limit,
in the calculations of this article.\\

%%%%%%%%%%%%%%%%%%%%%%%%%%%%%%%%%%%%%%%%%%%%%%%%%%%

%%%%%%%%%%%%%%%%%%%%%%%%%%
%%% syuusei 3
%%%%%%%%%%%%%%%%%%%%%%%%%%
\section{Normalization of the Partition Function} \label{norm}

In this appendix, we give the precise definition of the path integral measure
to decide the partition function without ambiguity.

As mentioned in section \ref{randr},
the absolute value of 
the infinite dimensional integrals of fluctuations around each vacuum
should be normalized to be $1$.
This is implemented by virtue of the supersymmetry.

When we normalize fields appropriately
the action of topological field theory
has the following form,
\begin{equation}
S_{TFT} = \int {\hat \delta}_+ [ \chi_i \{ H_i - i M_{i j} A_i \} ],
\label{stft}
\end{equation}
here we have omitted terms including fields like $\phi, {\bar \phi},
\eta$, often called ``Higgs sector'', for simplicity.
The normalization of the Higgs sector is possible to be managed
similarly to other fields when usual gauge fixing is done 
by using Nakanishi-Lautrup field, ghost and anti-ghost fields.
We can see this fact in the latter half of this section devoted to
trace and extra parts.
Also we have kept only quadratic terms of fluctuations, because
the path integral of topological field theories is estimated exactly in
the weak coupling limit.
$M_{i j}$ in (\ref{stft}) depends on backgrounds and
parameters in general,
but as seen below, the $M_{i j}$-dependence does not appear in the
result up to sign.
The BRS transformation rules are given by
\begin{eqnarray}
{\hat \delta}_+ A_i = \psi_i & , & {\hat \delta}_+ \psi_i = 0, \nonumber
 \\
{\hat \delta}_+ \chi_i = H_i & , & {\hat \delta}_+ H_i = 0.
\end{eqnarray}
For $A_i, \cdots$, we adopt the following path integral measure,
\begin{equation}
\prod_i \frac{d H_i}{\sqrt{2 \pi}} \frac{d A_i}{\sqrt{2 \pi}} d \chi_i d \psi_i,
\end{equation}
then we obtain
\begin{equation}
\lvert \int \prod_i \frac{d H_i}{\sqrt{2 \pi}} \frac{d A_i}{\sqrt{2 \pi}} d \chi_i d
 \psi_i e^{-S_{TFT}} \rvert = 1 .
\label{intstft}
\end{equation}
The $M_{i j}$-dependence does not appear due
to the supersymmetry.

Now we give a detailed argument for calculations about the trace and
 extra parts of our model as an example.
The action including the trace and extra parts $S^{\infty}_{tr \oplus
ex} + S_{g.f.} $ is decomposed into two parts, $S_1$ and $S_2$. $S_1$ consists of
(\ref{b1})-(\ref{b4}),(\ref{f1})-(\ref{f4}), and also $S_2$ consists of (\ref{b5}),(\ref{f5}),(\ref{gf1}),(\ref{gf2}).
$S_2$ involves the Higgs sector and also includes the gauge fixing
 terms.
$S_1$ involves all the rest.

We start with the $S_1$ part.
$S_1$ is represented in the same form as (\ref{stft}),
therefore we obtain
\begin{equation}
\int \prod_i \frac{d H_i}{\sqrt{2 \pi}} \frac{d A_i}{\sqrt{2 \pi}} d \chi_i d
 \psi_i e^{-S_1} = 1 .
\label{ints1}
\end{equation}
As mentioned above, the $\theta$-dependence does not appear.

Let us turn to the $S_2$ part.
The action is given as
\begin{eqnarray}
S_2 &=& \frac{4}{\theta} {\bar \phi}_{({\bf 1})} \phi_{({\bf 1})}
 \label{s2b1} \\
 &+& \frac{i}{\sqrt{\theta}} \eta_{({\bf 1})} \psi^{\mu}_{(\mu)}
  \label{s2f1} \\
 &+& b_{({\bf 1})} \left( b_{({\bf 1})} - \frac{1}{\sqrt{\theta}}
		      A_{\mu (\mu)} \right) \label{s2b2} \\
 &+& \frac{4}{\theta} {\bar \rho}_{({\bf 1})} \rho_{({\bf 1})} +
  \frac{1}{\sqrt{\theta}} {\bar \rho}_{({\bf 1})} \psi_{\mu (\mu)}. \label{s2f2}
\end{eqnarray}
We adopt the following measure,
\begin{equation}
\frac{d {\bar \phi}_{({\bf 1})}}{\sqrt{2 \pi}} \frac{d
 \phi_{({\bf 1})}}{\sqrt{2 \pi}} d {\bar \rho}_{({\bf 1})} d \rho_{({\bf
 1})} \frac{d b_{({\bf 1})}}{\sqrt{2 \pi}} \frac{d A_{\mu
 (\mu)}}{\sqrt{2 \pi}} d \eta_{({\bf 1})} d \psi^{\mu}_{(\mu)} ,
\end{equation}
then we obtain
\begin{equation}
\int \frac{d {\bar \phi}_{({\bf 1})}}{\sqrt{2 \pi}} \frac{d
 \phi_{({\bf 1})}}{\sqrt{2 \pi}} d {\bar \rho}_{({\bf 1})} d \rho_{({\bf
 1})} \frac{d b_{({\bf 1})}}{\sqrt{2 \pi}} \frac{d A_{\mu
 (\mu)}}{\sqrt{2 \pi}} d \eta_{({\bf 1})} d \psi^{\mu}_{(\mu)} e^{-S_2} =  1.
\label{ints2}
\end{equation}
Notice that the result (\ref{ints2}) is again a consequence of the
supersymmetry.

As a result of these normalizations, partition functions of the 
cohomological field theories are defined as well-defined functions
or finite values without ambiguity from infinite dimensional integral.

At the end of this appendix,
we should notice a fact relating the dimension-independence of partition function, (\ref{d=8}).
The gauge symmetry (\ref{gauge}) and the gauge fixing term
(\ref{strexgf0}) are expected to have the same form for all cases of (8-dim , ${\mathcal N}=2$),(6-dim , ${\mathcal N}=2$),
(4-dim , ${\mathcal N}=4$) and (2-dim , ${\mathcal N}=8$).
So we expect that the trace and extra sector produce a trivial factor $1$ for all of those cases.

%%%%%%%%%%%%%%%%%%%%%%%%%%%%%%%%%%%%%%%%%%%%%%%%
%%%%%%%%%%%%%%%%%%%%%%%%%%%%%%%%%%%%%%%%%%%%%%%%
%%%%%%%%%%%%%%%%%%%%%%%%%%%%%%%%%%%%%%%%%%%%%%%%

\end{document}